\documentclass{article}
\usepackage{amssymb}
\usepackage{latexsym}
\usepackage{graphicx}

\newcommand{\ket}[1]{\left | \, #1 \right \rangle}
\newcommand{\bra}[1]{\left \langle #1 \, \right |}

\begin{document}

\title{Basic concepts in quantum computation}
\author{Artur Ekert, Patrick Hayden and Hitoshi Inamori \\
Centre for Quantum Computation,\\
University of Oxford, Oxford OX1 3PU, United Kingdom
}
\maketitle

\section{Qubits, gates and networks}
Consider the two binary strings,
\begin{eqnarray}
&&011, \\
&&111.
\end{eqnarray}
The first one can represent, for example, the number $3$ (in binary)
and the second one the number $7.$ In general three physical bits can
be prepared in $2^{3}=8$ different configurations that can
represent, for example, the integers from $0$ to $7.$ However, a
register composed of three classical bits can store only one
number at a given moment of time. Enter qubits and quantum
registers:

A \emph{qubit} is a quantum system in which the Boolean states $0$
and $1$ are represented by a prescribed pair of normalised and
mutually orthogonal quantum states labeled as $\{|0\rangle
,|1\rangle \}$~\cite{Sch95}. The two states form a `computational
basis' and any other (pure) state of the qubit can be written as a
superposition $\alpha |0\rangle +\beta |1\rangle $ for some $\alpha
$ and $\beta $ such that $|\alpha |^{2}+|\beta |^{2}=1.$ A qubit is
typically a microscopic system, such as an atom, a nuclear spin, or
a polarised photon. A collection of $n$ qubits is called a
\emph{quantum register} of size $n$.

We shall assume that information is stored in the registers in
binary form. For example, the number $6$ is represented by a register
in state $|1\rangle \otimes |1\rangle \otimes |0\rangle $. In more
compact notation: $|a\rangle $ stands for the tensor product
$|a_{n-1}\rangle \otimes |a_{n-2}\rangle \ldots |a_{1}\rangle
\otimes |a_{0}\rangle $, where $a_i\in\{0,1\}$, and it represents
a quantum register prepared with the value
$a=2^{0}a_{0}+2^{1}a_{1}+\ldots2^{n-1}a_{n-1}$. There are $2^n$
states of this kind, representing all binary strings of length $n$
or numbers from $0$ to $2^n-1$, and they form a convenient
computational basis. In the following $a\in \{0,1\}^n$ ($a$ is a
binary string of length $n$) implies that $\ket{a}$ belongs to the
computational basis.

Thus a quantum register of size three can store individual numbers
such as $3$ or $7$,
\begin{eqnarray}
|0\rangle \otimes |1\rangle \otimes |1\rangle &\equiv &|011\rangle \equiv
|3\rangle , \\
|1\rangle \otimes |1\rangle \otimes |1\rangle &\equiv &|111\rangle \equiv
|7\rangle ,
\end{eqnarray}
but, it can also store the two of them simultaneously. For if we
take the first qubit and instead of setting it to $|0\rangle $ or
$|1\rangle $ we prepare a superposition $1/\sqrt{2}\left( |0\rangle
+|1\rangle \right)$ then we obtain
\begin{eqnarray}
\frac{1}{\sqrt{2}}\left( |0\rangle +|1\rangle \right) \otimes |1\rangle
\otimes |1\rangle &\equiv &\frac{1}{\sqrt{2}}\left( |011\rangle +|111\rangle
\right) , \\
&\equiv &\frac{1}{\sqrt{2}}\left( |3\rangle +|7\rangle \right) .
\end{eqnarray}
In fact we can prepare this register in a superposition of all
eight numbers -- it is enough to put each qubit into the
superposition $1/\sqrt{2}
\left( |0\rangle +|1\rangle \right) .$ This gives
\begin{equation}
\frac{1}{\sqrt{2}}\left( |0\rangle +|1\rangle \right) \otimes \frac{1}{\sqrt{
2}}\left( |0\rangle +|1\rangle \right) \otimes \frac{1}{\sqrt{2}}\left(
|0\rangle +|1\rangle \right) ,
\end{equation}
which can also be written in binary as (ignoring the normalisation
constant $ 2^{-3/2}$ ),
\begin{equation}
|000\rangle +|001\rangle +|010\rangle +|011\rangle +|100\rangle +|101\rangle
+|110\rangle +|111\rangle .
\end{equation}
or in decimal notation as
\begin{equation}
|0\rangle +|1\rangle +|2\rangle +|3\rangle +|4\rangle +|5\rangle
+|6\rangle +|7\rangle ,
\end{equation}
or simply as
\begin{equation}
\sum_{x=0}^7\ket{x}.
\end{equation}

These preparations, and any other manipulations on qubits, have to
be performed by unitary operations. A \emph{quantum logic gate} is
a device which performs a fixed unitary operation on selected
qubits in a fixed period of time and a \emph{quantum network} is a
device consisting of quantum logic gates whose computational steps
are synchronised in time~\cite{Deu89}. The outputs of some of the
gates are connected by wires to the inputs of others. The
\emph{size} of the network is the number of gates it contains.

The most common quantum gate is the Hadamard gate, a single
qubit gate $H$ performing the unitary transformation known as the
Hadamard transform.  It is defined as
\setlength{\unitlength}{0.030in}
\[
H=\frac{1}{\sqrt{2}}\left(
\begin{array}{cc}
1 & 1 \\
1 & -1
\end{array}
\right) \mbox{\hspace{2cm}}
\mbox{
\begin{picture}(30,0)(15,15)
  \put(-4,14){$\ket{x}$} \put(5,15){\line(1,0){5}} \put(20,15){\line(1,0){5}}
  \put(10,10){\framebox(10,10){$H$}}
\put(30,14){$(-1)^x\ket{x}+\ket{1-x}$}
\end{picture}
}\nonumber
\]
The matrix is written in the computational basis
$\{\ket{0}, \ket{1} \}$
and the diagram on
the right provides a schematic representation of the gate $H$
acting on a qubit in state $\ket{x}$, with $x=0,1$.

And here is a network, of size three, which affects the Hadamard
transform on three qubits. If they are initially in state
$\ket{000}$ then the output is the superposition of all eight
numbers from $0$ to $7$. \\

\setlength{\unitlength}{0.025 in}
\begin{center}
\begin{picture}(160,40)
\put(0,33){$\ket{0}$}
\put(10,35){\line(1,0){10}}
\put(20,30){\framebox(10,10){H}}
\put(30,35){\line(1,0){10}}
\put(45,33){$\frac{\ket{0}+\ket{1}}{\sqrt{2}}$}

\put(0,18){$\ket{0}$}
\put(10,20){\line(1,0){10}}
\put(20,15){\framebox(10,10){H}}
\put(30,20){\line(1,0){10}}
\put(45,18){$\frac{\ket{0}+\ket{1}}{\sqrt{2}}$}

\put(0,3){$\ket{0}$}
\put(10,5){\line(1,0){10}}
\put(20,0){\framebox(10,10){H}}
\put(30,5){\line(1,0){10}}
\put(45,3){$\frac{\ket{0}+\ket{1}}{\sqrt{2}}$}

\put(60,20){$\left.\begin{array}{c}\\ \\ \\ \\ \\ \\ \\ \\ \end{array}\right\}$}
\put(75,20){\scriptsize
$
\begin{array}{l}
\quad\mbox{IN BINARY}\\
\\
=\frac{1}{2^{3/2}}\left\{ \begin{array}{l} \ket{000}+\ket{001}+\ket{010}
+\ket{011}+\\+\ket{100}+\ket{101}+\ket{110}+\ket{111}\end{array}\right\} \\
\\
=\frac{1}{2^{3/2}}\left\{\begin{array}{l}
\ket{0}+\ket{1}+\ket{2}+\ket{3}+ \\ +\ket{4}+\ket{5}+\ket{6}+\ket{7}
\end{array}\right\}\\
\\
\quad\mbox{IN DECIMAL}
\end{array}$}
\end{picture}
\end{center}

If the three qubits are initially in some other state from the
computational basis then the result is a superposition of all
numbers from $0$ to $7$ but exactly half of them will appear in the
superposition with the minus sign, for example,
\begin{equation}
\ket{101}\mapsto\frac{1}{2^{3/2}}\left\{\begin{array}{l} \ket{000}
-\ket{001}+\ket{010}
-\ket{011}+\\-\ket{100}+\ket{101}-\ket{110}+\ket{111}\end{array}\right\}.
\end{equation}
In general, if we start with a register of size $n$ in some state
$y\in\{0,1\}^n$ then
\begin{equation}
\ket{y}\mapsto 2^{-n/2} \sum_{x\in\{0, 1\}^n}(-1)^{y\cdot x}\ket{x},
\label{Had}
\end{equation}
where the product of $y =(y_{n-1},\ldots , y_0)$ and
$x=(x_{n-1},\ldots, x_0)$ is taken bit by bit:
\begin{equation}
y\cdot x = (y_{n-1}x_{n-1}+\ldots y_1x_1 +y_0x_0).
\label{binprod}
\end{equation}

We will need another single qubit gate -- the phase shift gate
$\mathbf{\phi }$ defined as $\left|
\,0\right\rangle \mapsto \left| \,0\right\rangle $ and $\left|
\,1\right\rangle \mapsto e^{i\phi }\left| \,1\right\rangle $, or,
in matrix notation,
\setlength{\unitlength}{0.030in}
\begin{equation}
{\mathbf{\phi}} = \left (
\begin{array}{cc}
1 & 0 \\
0 & e^{i\phi}
\end{array}
\right ) \mbox{\hspace{3cm}}
\mbox{
\begin{picture}(30,0)(15,15)
  \put(-4,14){$\ket{x}$} \put(5,15){\line(1,0){20}} \put(20,15){\line(1,0){5}}
  \put(15,15){\circle*{3}}
\put(14,19){$\phi $}
\put(30,14){$e^{ix\phi}\ket{x}$}
\end{picture}
}
\end{equation}

The Hadamard gate and the phase gate can be combined to construct
the following network (of size four), which generates the most
general pure state of a single qubit (up to a global phase),
\setlength{\unitlength}{0.025in}
\begin{equation}\label{1quniv}
\mbox{
\begin{picture}(80,20)(0,3)

\put(0,5){$\ket{0}$}

\put(10,5){\line(1,0){10}}
\put(20,0){\framebox(10,10){H}}
\put(30,5){\line(1,0){20}}
\put(50,0){\framebox(10,10){H}}
\put(60,5){\line(1,0){20}}

\put(40,5){\circle*{3}}
\put(38,10){$2\theta$}

\put(70,5){\circle*{3}}
\put(65,10){$\frac{\pi}{2}+\phi$}

\end{picture}
}
\quad
\cos\theta\ket{0}+e^{i\phi}\sin\theta\ket{1}.
\end{equation}
Consequently, the Hadamard and phase gates are sufficient to construct
\emph{any} unitary operation on a single qubit.

Thus the Hadamard gates and the phase gates can be used to
transform the input state $|0\rangle |0\rangle ...|0\rangle $ of
the $n$ qubit register into any state of the type $|\Psi
_{1}\rangle $ $|\Psi _{2}\rangle ...$ $ |\Psi _{n}\rangle ,$ where
$|\Psi _{i}\rangle $ is an arbitrary superposition of $|0\rangle $
and $|1\rangle .$ These are rather special $n$-qubit states, called
the product states or the separable states. In general, a quantum
register of size $n>1$ can be prepared in states which are not
separable -- they are known as entangled states. For example, for
two qubits ($n=2$), the state
\begin{equation}
\alpha \ |00\rangle +\beta \ |01\rangle =|0\rangle \otimes\left( \alpha \ |0\rangle
+\beta \ |1\rangle \right)
\end{equation}
is separable, $\ket{\Psi_1}=|0\rangle$ and $\ket{\Psi_2}= \alpha \
|0\rangle +\beta \ |1\rangle$, whilst the state
\begin{equation}
\alpha \ |00\rangle +\beta \ |11\rangle\neq
\ket{\Psi_1}\otimes\ket{\Psi_2}
\end{equation}
is entangled ($\alpha ,\beta \neq 0$), because it cannot be written
as a tensor product.

In order to entangle two (or more qubits) we have to extend our
repertoire of quantum gates to two-qubit gates. The most popular
two-qubit gate is the controlled-NOT (\textsc{c-not}), also known
as the \textsc{xor} or the measurement gate. It flips the second
(target) qubit if the first (control) qubit is $\left|
\,1\right\rangle $ and does nothing if the control qubit is $\left|
\,0\right\rangle $. The gate is represented by the unitary matrix
\begin{equation}
C=\left(
\begin{array}{cccc}
1 & 0 & 0 & 0 \\
0 & 1 & 0 & 0 \\
0 & 0 & 0 & 1 \\
0 & 0 & 1 & 0
\end{array}
\right) \mbox{\hspace{1.5cm}}
\mbox{
\begin{picture}(25,0)(0,20)
  \put(-4,14){$\ket{y}$} \put(-4,29){$\ket{x}$} \put(5,15){\line(1,0){20}}
\put(5,30){\line(1,0){20}}
  \put(15,30){\circle*{3}} \put(15,11){\line(0,1){19}}
\put(15,15){\circle{8}}
\put(27,14){$\ket{x\oplus y}$}
\put(27,29){$\ket{x}$}
\end{picture}}
\end{equation}
where $x,y=0\mbox{ or }1$ and $\oplus $ denotes XOR or addition modulo 2. If
we apply the \textsc{c-not} to Boolean data in which the target qubit
is $|0\rangle $
and the control is either $|0\rangle $ or $|1\rangle $ then the effect is to
leave the control unchanged while the target becomes a copy of the control,
\emph{i.e.}
\begin{equation}
|x\rangle |0\rangle \mapsto |x\rangle |x\rangle \qquad x=0,1.
\end{equation}
One might suppose that this gate could also be used to copy superpositions
such as $|\Psi \rangle =\alpha \ |0\rangle +\beta \ |1\rangle ,$ so that
\begin{equation}
|\Psi \rangle |0\rangle \mapsto |\Psi \rangle |\Psi \rangle
\label{cloning}
\end{equation}
for any $|\Psi \rangle .$ This is not so! The unitarity of the \textsc{c-not}
requires that the gate turns superpositions in the control qubit into\emph{\
entanglement} of the control and the target. If the control qubit is in a
superposition state $|\Psi \rangle =\alpha |0\rangle +\beta |1\rangle ,$
\noindent $(\alpha ,\beta \neq 0),$ and the target in $|0\rangle $ then the
\textsc{c-not} generates the entangled state
\begin{equation}
\left( \alpha |0\rangle +\beta |1\rangle \right) |0\rangle \mapsto
\alpha |00\rangle +\beta |11\rangle.
\end{equation}

Let us notice in passing that it is impossible to construct a
universal quantum cloning machine effecting
the transformation in Eq.(\ref{cloning}), or even the more general
\begin{equation}
|\Psi \rangle |0\rangle |W\rangle \mapsto |\Psi \rangle |\Psi \rangle
|W^{\prime }\rangle
\end{equation}
where $|W\rangle $ refers to the state of the rest of the world and $|\Psi
\rangle $ is \emph{any} quantum state~\cite{WZ82}. To see this take any
two normalised states $|\Psi \rangle $ and $|\Phi \rangle $ which are
non-identical ($|\langle \Phi |\Psi \rangle| \neq 1)$ and non-orthogonal ($
\langle \Phi |\Psi \rangle \neq 0$ ), and run the cloning machine,
\begin{eqnarray}
|\Psi \rangle |0\rangle |W\rangle &\mapsto &|\Psi \rangle |\Psi \rangle
|W^{\prime }\rangle \\
|\Phi \rangle |0\rangle |W\rangle &\mapsto &|\Phi \rangle |\Phi \rangle
|W^{^{\prime \prime }}\rangle
\end{eqnarray}
As this must be a unitary transformation which preserves the inner
product hence we must require
\begin{equation}
\langle \Phi |\Psi \rangle =\langle \Phi |\Psi \rangle ^{2}\langle
W^{^{\prime }}|W^{^{\prime \prime }}\rangle
\end{equation}
and this can only be satisfied when $|\langle \Phi |\Psi \rangle| =0$
or $1$, which contradicts our assumptions. Thus states of qubits,
unlike states of classical bits, cannot be faithfully cloned. This
leads to interesting applications, quantum cryptography being one
such.

Another common two-qubit gate is the controlled phase shift gate
$B(\phi)$ defined as
\begin{equation}
{B}(\phi )=\left. \left(
\begin{array}{cccc}
1 & 0 & 0 & 0 \\
0 & 1 & 0 & 0 \\
0 & 0 & 1 & 0 \\
0 & 0 & 0 & e^{i\phi }
\end{array}
\right) \mbox{\hspace{1.5cm}}
\mbox{
\begin{picture}(25,0)(0,20)
  \put(-4,14){$\ket{y}$} \put(-4,29){$\ket{x}$} \put(5,15){\line(1,0){20}}
\put(5,30){\line(1,0){20}}
  \put(15,30){\circle*{3}} \put(15,15){\line(0,1){15}}
\put(15,15){\circle*{3}}
\end{picture}}\quad \right\} e^{ixy\phi }\left| \,x\right\rangle \left|
\,y\right\rangle .
\end{equation}
Again, the matrix is written in the computational basis
$\{ \ket{00}, \ket{01}, \ket{10},$ $\ket{11} \}$
and the diagram on the right shows the structure of the gate.

More generally, these various 2-qubit controlled gates are all
of the form controlled-$U$, for some single-qubit unitary
transformation $U$.
The controlled-$U$ gate applies the identity transformation to the auxiliary
(lower) qubit when the control qubit is in state
$\ket{0}$ and applies an arbitrary prescribed $U$ when the
control qubit is in state $\ket{1}$. The gate maps
$\ket{0}\ket{y}$ to $\ket{0}\ket{y}$
and $\ket{1}\ket{y}$ to $\ket{1}(U \ket{y})$,
and is graphically represented as
\setlength{\unitlength}{0.033in}
\begin{center}
\begin{picture}(30,30)(0,10)
\put(0,15){\line(1,0){10}}
\put(20,15){\line(1,0){10}}
\put(0,30){\line(1,0){30}}
\put(15,30){\circle*{3}}
\put(15,20){\line(0,1){10}}
\put(10,10){\framebox(10,10){$U$}}
\end{picture}
\end{center}

The Hadamard gate, all phase gates, and the \textsc{c-not}, form an
infinite {\em universal set of gates}
\emph{i.e.} if the \textsc{c-not} gate as well as the
Hadamard and all phase gates are available then any $n$-qubit
unitary operation can be simulated exactly with $O(4^{n}n)$ such
gates~\cite{BBC95}. (Here and in the following we use the asymptotic
notation -- $O(T(n))$ means bounded above by $c\,T(n)$ for some
constant $c>0$ for sufficiently large $n$.) This is not the only
universal set of gates. In fact, almost any gate which can entangle
two qubits can be used as a universal gate~\cite{BDEJ95,Llo95}.
Mathematically, an elegant choice is a pair of the Hadamard and the
controlled-$V$
\noindent (\textsc{c}-$V$) where $V$ is described by the unitary matrix
\begin{equation}
V=\left(
\begin{array}{cc}
1 & 0 \\
0 & i
\end{array}
\right) .
\end{equation}
The two gates form a finite universal set of gates --
networks containing only a finite number of these gates can
approximate any unitary transformation on two
(and more) qubits. More precisely, if $U$ is any two-qubit gate and
$\varepsilon >0$ then there exists a quantum network of size
$O(\log ^{d}(1/\varepsilon ))$ (where $d$ is a constant) consisting
of only $H$ and \textsc{c}-$V$ gates which computes a unitary operation
$U^{\prime }$ that is within distance $\varepsilon $ from
$U$~\cite{Sol99}. The metric is induced by the Euclidean norm
- we say that $U^{\prime }$ is within distance $\varepsilon $
from $U$ if there exists a unit complex number $\lambda $ (phase
factor) such that $||U-\lambda U^{\prime }||\leq \varepsilon $.
Thus if $U^{\prime }$ is substituted for $U$ in a quantum network
then the final state $\sum_{x}\alpha _{x}^{\prime }\left|
x\right\rangle $ approximates the final state of the original
network $
\sum_{x}\alpha _{x}\left| x\right\rangle $ as follows: $\sqrt{
\sum_{x}|\lambda \alpha _{x}^{\prime }-\alpha _{x}|^{2}}\leq \varepsilon $.
The probability of any specified measurement outcome on the final
state is affected by at most $\varepsilon $.

A \emph{quantum computer} will be viewed here as a quantum network
(or a family of quantum networks)and quantum computation is defined
as a unitary evolution of the network which takes its initial state
``input'' into some final state ``output''. We have chosen the
network model of computation, rather than Turing machines, because
it is relatively simple and easy to work with and because it is
much more relevant when it comes to physical implementation of
quantum computation.

\section{Quantum arithmetic and function evaluations}
Let us now describe how quantum computers actually compute, how
they add and multiply numbers, and how they evaluate Boolean
functions by means of unitary operations. Here and in the following
we will often use the modular arithmetic~\cite{HW79}. Recall that
\begin{equation}
a\bmod{b}
\end{equation}
denotes the remainder obtained by dividing integer $b$ into integer
$a$, which is always a number less than $b$. Basically $a=b\bmod n$
if $a=b+kn$ for some integer $k$. This is expressed by saying that
$a$ is \emph{congruent} to $b$ modulo $n$ or that $b$ is the\emph{\
residue} of $a$ modulo $n$. For example, $1\bmod 7=8\bmod 7=15\bmod
7=50\bmod 7=1$. Modular arithmetic is commutative, associative, and
distributive.
\begin{eqnarray}
(a\pm b)\bmod n &=&((a\bmod n)\pm (b\bmod n))\bmod n \\ (a\times
b)\bmod n &=&((a\bmod n)\times (b\bmod n))\bmod n \\ (a\times
(b+c))\bmod n &=&(((a b)\bmod n+((a c)\bmod n))\bmod n
\end{eqnarray}
Thus, if you need to calculate, say, $3^8\bmod 7$ do not use the
naive approach and perform seven multiplications and one huge
modular reduction. Instead, perform three smaller multiplications
and three smaller reductions,
\begin{equation}
((3^2\bmod 7)^2\bmod 7)^2\bmod 7 = (2^2\bmod 7)^2\bmod 7=16 \bmod
7= 2.
\end{equation}
This kind of arithmetic is ideal for computers as it restricts the
range of all intermediate results. For $l$-bit modulus $n$, the
intermediate results of any addition, subtraction or multiplication
will not be more than $2l$ bits long. In quantum registers of size
$n$, addition modulo $2^n$ is one of the most common operations; for
all $x \in \{0,1\}^n$ and for any $a\in \{0,1\}^n$,
\begin{equation}
\ket{x}\mapsto\ket{(x+a)\bmod 2^n}
\end{equation}
is a well defined unitary transformation.

The tricky bit in the modular arithmetic is the inverse operation,
and here we need some basic number theory. An integer $a\geq 2$ is
said to be \emph{prime} if it is divisible only by 1 and $a$ (we
consider only positive divisors). Otherwise, $a$ is called
\emph{composite}. The greatest common divisor of two integers $a$
and $b$ is the greatest positive integer $d$ denoted $d=\gcd(a,b)$
that divides both $a$ and $b$. Two integers $a$ and $b$ are said to
be \emph{coprime} or \emph{relatively prime} if $\gcd (a,b)=1$.
Given two integers $a$ and $n$ that are coprime, it can be shown
that there exists an unique integer $d\in\{0,\ldots,n-1\}$ such
that $a d=1\bmod n$~\cite{HW79}. The integer $d$ is called
\emph{inverse modulo } $n$ of $a$, and denoted $a^{-1}$. For
example, modulo $7$ we find that $3^{-1}=5
\bmod n$, since $3
\times 5 = 15 = 2\times 7 +1=1 \bmod 7$. This bizarre arithmetic
and the notation is due to Karl Friedrich Gauss (1777-1855). It was
first introduced in his \textit{ Disquistiones Arithmeticae }in
1801.

In quantum computers addition, multiplication, and any other
arithmetic operation have to be embedded in unitary evolution. We
will stick to the Hadamard and the controlled-$V$ (\textsc{c}-$V$), and use
them as building blocks for all other gates and eventually for
quantum adders and multipliers.

If we apply \textsc{c}-$V$ four times we get identity, so any three
subsequent applications of \textsc{c}-$V$ give the inverse of
\textsc{c}-$V$, which
will be called \textsc{c}-$V^{\dagger }$. Now, if we have a couple of the
\textsc{c}-$V$ gates and a couple of the Hadamard gates we can build the
\textsc{c-not} as follows
\setlength{\unitlength}{0.0308in}
\begin{center}
\begin{picture}(105, 40)
\put(0,10){\line(1,0){10}}
\put(20,10){\line(1,0){5}}
\put(35,10){\line(1,0){5}}
\put(50,10){\line(1,0){5}}
\put(65,10){\line(1,0){10}}

\put(0,30){\line(1,0){75}}

\put(10,5){\framebox(10,10){$H$}}
\put(25,5){\framebox(10,10){$V$}}
\put(40,5){\framebox(10,10){$V$}}
\put(55,5){\framebox(10,10){$H$}}

\put(30,30){\circle*{3}}
\put(30,30){\line(0,-1){15}}

\put(45,30){\circle*{3}}
\put(45,30){\line(0,-1){15}}

\put(80,15){\makebox(10,10){$\equiv$}}

\put(90,10){\line(1,0){15}}
\put(90,30){\line(1,0){15}}
\put(97.5,30){\circle*{3}}
\put(97.5,30){\line(0,-1){23}}
\put(97.5,10){\circle{6}}

\end{picture}
\end{center}

A single qubit operation \textsc{not} can be performed via a
\textsc{c-not} gate if
the control qubit is set to $|1\rangle $ and viewed as an auxiliary
qubit. This is not to say that we want to do it in practice. The
\textsc{c-not} gate is much more difficult to build than a single qubit
\textsc{not}.
Right now we are looking into the mathematical structure of quantum
Boolean networks and do not care about practicalities. Our two
elementary gates also allow us to construct a very useful gate called
the controlled-controlled-\textsc{not} gate ($c^{2}$-\textsc{not})
or the Toffoli
gate~\cite{Tof81}. The construction is given by the following
network,
\setlength{\unitlength}{0.025in}
\begin{center}
\begin{picture}(120,45)(0,10)

\put(0,15){\line(1,0){10}}
\put(20,15){\line(1,0){5}}
\put(35,15){\line(1,0){20}}
\put(65,15){\line(1,0){20}}
\put(95,15){\line(1,0){5}}
\put(110,15){\line(1,0){10}}

\put(0,30){\line(1,0){120}}
\put(0,45){\line(1,0){120}}

\put(45,30){\circle{6}}
\put(75,30){\circle{6}}

\put(30,30){\circle*{3}}
\put(60,30){\circle*{3}}
\put(45,45){\circle*{3}}
\put(75,45){\circle*{3}}
\put(90,45){\circle*{3}}

\put(90,20){\line(0,1){25}}
\put(30,20){\line(0,1){10}}
\put(45,27){\line(0,1){18}}
\put(60,20){\line(0,1){10}}
\put(75,27){\line(0,1){18}}

\put(10,10){\framebox(10,10){$H$}}
\put(25,10){\framebox(10,10){$V$}}
\put(55,10){\framebox(10,10){$V^{\mbox{\scriptsize \dag}}$}}
\put(85,10){\framebox(10,10){$V$}}
\put(100,10){\framebox(10,10){$H$}}

\end{picture}
\begin{picture}(14,45)(0,2.5)

\put(0,0){\makebox(14,45){$\equiv$}}

\end{picture}
\begin{picture}(22,45)(0,10)

\put(0,15){\line(1,0){22}}
\put(0,30){\line(1,0){22}}
\put(0,45){\line(1,0){22}}

\put(11,30){\circle*{3}}
\put(11,45){\circle*{3}}

\put(11,12){\line(0,1){33}}

\put(11,15){\circle{6}}

\end{picture}
\end{center}
This gate has two control qubits (the top two wires on the diagram)
and one target qubit which is negated only when the two controls
are in the state $ |1\rangle |1\rangle $. The $c^2$-\textsc{not} gate gives
us the logical connectives we need for arithmetic. If the target is
initially set to $|0\rangle $ the gate acts as a reversible \textsc{and}
gate - after the gate operation the target becomes the logical \textsc{and}
of the two control qubits.
\begin{equation}
\ket{x_1,x_2}\ket{0}\mapsto \ket{x_1,x_2}\ket{x_1\wedge
x_2}
\label{andgate}
\end{equation}
Once we have in our repertoire operations such as \textsc{not},
\textsc{and}, and
\textsc{c-not}, all of them implemented as unitary operations, we can, at
least in principle, evaluate any Boolean function
$\{0,1\}^{n}\rightarrow \{0,1\}^{m}$ which map $n$ bits of input
into $m$ bits of output. A simple concatenation of the Toffoli gate
and the \textsc{c-not} gives a simplified quantum adder, shown below, which
is a good starting point for constructing full adders, multipliers
and more elaborate networks.
\setlength{\unitlength}{0.015in}
\begin{center}
\begin{picture}(180,80)(110,0)

\put(70,56){\makebox(20,12){$\ket{x_1}$}}
\put(70,36){\makebox(20,12){$\ket{x_2}$}}
\put(70,16){\makebox(20,12){$\ket{y}$}}
\put(90,60){\line(1,0){40}}
\put(90,40){\line(1,0){40}}
\put(90,20){\line(1,0){40}}

\put(110,60){\circle*{4}}
\put(110,40){\circle*{4}}
\put(110,20){\circle{8}}
\put(110,60){\line(0,-1){44}}

\put(130,56){\makebox(20,12){$\ket{x_1}$}}
\put(130,36){\makebox(20,12){$\ket{x_2}$}}
\put(139,16){\makebox(20,12){$\ket{x_1x_2\oplus y}$}}

\put(195,60){$\ket{x_1}$}
\put(195,40){$\ket{x_2}$}
\put(195,20){$\ket{0}$}
\put(275,60){$\ket{x_1}$}
\put(275,40){SUM = $\ket{x_1\oplus x_2}$}
\put(275,20){CARRY = $\ket{x_1x_2}$}

\put(210,60){\line(1,0){60}}
\put(210,40){\line(1,0){60}}
\put(210,20){\line(1,0){60}}

\put(230,60){\circle*{4}}
\put(230,40){\circle*{4}}
\put(230,20){\circle{8}}
\put(230,60){\line(0,-1){44}}

\put(250,60){\circle*{4}}
\put(250,40){\circle{8}}
\put(250,60){\line(0,-1){24}}

\put(84,1){TOFFOLI GATE}
\put(208,1){QUANTUM ADDER}

\end{picture}
\end{center}
We can view the Toffoli gate and the evolution given by
Eq.~(\ref{andgate}) as a quantum implementation of a Boolean
function $f:\{0,1\}^{2}\rightarrow\{0,1\}$ defined by
$f(x_1,x_2) = x_1 \wedge x_2$.
The operation
\textsc{and} is not reversible, so we had to embed it in the reversible
operation $c^2$-\textsc{not}. If the third bit is initially set to $1$
rather than $0$ then the value of $x_1\wedge x_2$ is negated. In
general we write the action of the Toffoli gate as the function
evaluation,
\begin{equation}
\ket{x_1,x_2}\ket{y}\mapsto\ket{x_1,x_2}\ket{(y+(x_1\wedge x_2))\bmod
2}.
\end{equation}

This is how we compute any Boolean function
$\{0,1\}^{n}\rightarrow\{0,1\}^{m}$ on a quantum computer. We
require at least two quantum registers; the first one, of size $n$,
to store the arguments of $f$ and the second one, of size $n$, to
store the values of $f$. The function evaluation is then a unitary
evolution of the two registers,
\begin{equation}
|x,y\rangle \mapsto |x,(y+f(x))\bmod 2^{m}\rangle .
\end{equation}
for any $y\in\{0,1\}^m$. (In the following, if there is no danger of
confusion, we may simplify the notation and omit the $\bmod$
suffix.)

For example, a network computing $f:\{0,1\}^{2}\rightarrow
\{0,1\}^{3}$ such that $f(x)=x^{2}$ acts as follows
\begin{eqnarray}
|00\rangle |000\rangle &\mapsto &|00\rangle |000\rangle,\qquad
|10\rangle |000\rangle \mapsto |10\rangle |100\rangle \\
|01\rangle |000\rangle &\mapsto &|01\rangle |001\rangle,\qquad
|11\rangle |000\rangle \mapsto |11\rangle |001\rangle
\end{eqnarray}
which can be written as
\begin{equation}
|x,0\rangle \mapsto |x,x^{2}\bmod 8\rangle ,
\end{equation}
\emph{e.g.} $3^2\bmod 2^3=1$ which explains why $|11\rangle |000\rangle
\mapsto |11\rangle |001\rangle$.

In fact, for these kind of operations we also need a third register
with the so-called working bits which are set to zero at the input
and return to zero at the output but which can take non-zero values
during the computation.

What makes quantum function evaluation really interesting is its
action on a superposition of different inputs $x$, for example,
\begin{equation}
\sum_{x}|x,0\rangle \mapsto \sum_{x}|x,f(x)\rangle
\end{equation}
produces $f(x)$ for all $x$ in a single run. The snag is that we
cannot get them all from the entangled state
$\sum_{x}|x,f(x)\rangle $ because any bit by bit measurement on the
first register will yield one particular value
$x^{\prime}\in\{0,1\}^n$ and the second register will then be found
with the value $f(x^{\prime })\in\{0,1\}^m$.

\section{Algorithms and their complexity}
In order to solve a particular problem, computers, be it classical
or quantum, follow a precise set of instructions that can be
mechanically applied to yield the solution to any given instance of
the problem. A specification of this set of instructions is called
an algorithm. Examples of algorithms are the procedures taught in
elementary schools for adding and multiplying whole numbers; when
these procedures are mechanically applied, they always yield the
correct result for any pair of whole numbers. Any algorithm can be
represented by a family of Boolean networks $(N_1,N_2,N_3,... )$,
where the network $N_n$ acts on all possible input instances of
size $n$ bits. Any useful algorithm should have such a family
specified by an example network $N_n$ and {\em a simple rule}
explaining how to construct the network $N_{n+1}$ from the network
$N_{n}$. These are called {\em uniform} families of
networks~\cite{Pap94}.\footnote{This means that the network model
is not a self-contained model of computation. We need an algorithm,
a Turing machine, which maps each $n$ into an explicit description
of $N_n$.}

The quantum Hadamard transform defined by Eq.(\ref{Had}) has a
uniform family of networks whose size is growing as $n$ with the
number of input qubits. Another good example of a uniform family of
networks is the quantum Fourier transform (QFT)~\cite{Cop94}
defined in the computational basis as the unitary operation
\begin{equation} \label{Four}
\ket{y}\mapsto 2^{-n/2} \sum_x e^{i \frac{2\pi}{2^n} yx}\ket{x},
\end{equation}
Suppose we want to \emph{construct} such a unitary evolution of $n$
qubits using our repertoire of quantum logic gates. We can start
with a single qubit and notice that in this case the QFT is reduced
to applying a Hadamard gate. Then we can take two qubits and notice
that the QFT can be implemented with two Hadamard gates and the
controlled phase shift $B(\pi)$in between. Progressing this way we
can construct the three qubit QFT and the four qubit QFT, whose
network looks like this:
\setlength{\unitlength}{0.015in}
\begin{center}
\begin{picture}(210,120)(40,0)

\put(0,34){\makebox(20,12){$\ket{x_3}$}}
\put(0,54){\makebox(20,12){$\ket{x_2}$}}
\put(0,74){\makebox(20,12){$\ket{x_1}$}}
\put(0,94){\makebox(20,12){$\ket{x_0}$}}

\put(247,34){\makebox(20,12){$\ket{0}+e^{2\pi ix/2^4}\ket{1}$}}
\put(247,54){\makebox(20,12){$\ket{0}+e^{2\pi ix/2^3}\ket{1}$}}
\put(247,74){\makebox(20,12){$\ket{0}+e^{2\pi ix/2^2}\ket{1}$}}
\put(247,94){\makebox(20,12){$\ket{0}+e^{2\pi ix/2}\ket{1}$}}

\put(43,16){$H\;B(\pi)\;H\;B(\pi/2)B(\pi)\;H\;B(\pi/4)B(\pi/2)B(\pi)\;H$}

\put(20,40){\line(1,0){170}}
\put(190,34){\framebox(12,12){$H$}}
\put(202,40){\line(1,0){18}}
\put(20,60){\line(1,0){104}}
\put(124,54){\framebox(12,12){$H$}}
\put(136,60){\line(1,0){84}}
\put(20,80){\line(1,0){54}}
\put(74,74){\framebox(12,12){$H$}}
\put(86,80){\line(1,0){134}}
\put(20,100){\line(1,0){14}}
\put(34,94){\framebox(12,12){$H$}}
\put(46,100){\line(1,0){174}}

\put(60,100){\circle*{4}}
\put(60,100){\line(0,-1){20}}
\put(60,80){\circle*{4}}

\put(100,100){\circle*{4}}
\put(100,60){\circle*{4}}
\put(100,60){\line(0,1){40}}

\put(110,80){\circle*{4}}
\put(110,60){\circle*{4}}
\put(110,60){\line(0,1){20}}

\put(150,40){\circle*{4}}
\put(160,40){\circle*{4}}
\put(170,40){\circle*{4}}
\put(150,40){\line(0,1){60}}
\put(160,40){\line(0,1){40}}
\put(170,40){\line(0,1){20}}
\put(150,100){\circle*{4}}
\put(160,80){\circle*{4}}
\put(170,60){\circle*{4}}

\end{picture}
\end{center}
(\emph{N.B.} there are three different types of
the ${B}(\phi)$ gate in the network above: ${B}(\pi)$, $B(\pi/2)$
and $B(\pi/4)$.)

The general case of $n$ qubits requires a trivial extension of the
network following the same sequence pattern of gates $H$ and
$B$. The QFT network operating on $n$ qubits contains $n$
Hadamard gates $H$ and $n(n-1)/2$ phase shifts $B$, in
total $n(n+1)/2$ elementary gates.

The big issue in designing algorithms or their corresponding families of
networks is the optimal use of physical resources required to solve
a problem. Complexity theory is concerned with the inherent cost of
computation in terms of some designated elementary operations,
memory usage, or network size. An algorithm is said to be fast or
efficient if the number of elementary operations taken to execute
it increases no faster than a polynomial function of the size of
the input. We generally take the input size to be the total number
of bits needed to specify the input (for example, a number $N$
requires $\log_{2}N$ bits of binary storage in a computer). In the
language of network complexity - an algorithm is said to be
\emph{efficient} if it has a uniform and polynomial-size network
family ($O(n^d)$ for some constant $d$)~\cite{Pap94}. For example,
the quantum Fourier transform can be performed in an efficient way
because it has a uniform family of networks whose size grows only
as a quadratic function of the size of the input,
\emph{i.e.} $O(n^2)$.
Changing from one set of gates to another, \emph{e.g.} constructing the
QFT out of the Hadamard and the controlled-$V$ gates with a
prescribed precision $\epsilon$, can only affect the network size
by a multiplicative constant which does not affect the quadratic
scaling with $n$. Thus the complexity of the QFT is $O(n^2)$ no
matter which set of adequate gates we use. Problems which do not
have efficient algorithms are known as hard problems.

Elementary arithmetic operations taught at schools, such as long
addition, multiplication or division of $n$ bit numbers require
$O(n^2)$ operations. For example, to multiply
$x=(x_{n-1}...x_1x_0)$ and $y=(y_{n-1}...y_1y_0)$ we successively
multiply $y$ by $x_0$, $x_1$ and so on, shift, and then add the
result. Each multiplication of $y$ by $x_k$ takes about $n$ single
bit operations, the addition of the $n$ products takes of the order
of $n^2$ bit operations, which adds to the total $O(n^2)$
operations. Knowing the complexity of elementary arithmetic one can
often assess the complexity of other algorithms. For example, the
greatest common divisor of two integers $x$ and $y<x$
can be found using Euclid's algorithm; the oldest nontrivial
algorithm which has been known and used since 300 BC.\footnote{This
truly `classical' algorithm is described in Euclid's
\textit{Elements}, the oldest Greek treatise in mathematics to
reach us in its entirety. Knuth (1981) provides an extensive
discussion of various versions of Euclid's algorithm.} First divide
$x$ by $y$ obtaining remainder $r_1$. Then divide $y$ by $r_1$
obtaining remainder $r_2$, then divide $r_1$ by $r_2$ obtaining
remainder $r_3$, etc., until the remainder is zero. The last
non-zero remainder is $\gcd(x,y)$ because it divides all previous
remainders and hence also $x$ and $y$ (it is obvious from the
construction that it is the \emph{greatest} common divisor). For
example, here is a sequence or remainders $(r_j,r_{j+1})$ when we
apply Euclid's algorithm to compute $\gcd(12378,3054)=6$:
(12378,3054), (3054,162), (162, 138), (138, 24), (24, 18), (18,6),
(6,0). What is the complexity of this algorithm? It is easy to see
that the largest of the two numbers is at least halved every two
steps, so every two steps we need one bit less to represent the
number, and so the number of steps is at most $2n$, where $n$ is
the number of bits in the two integers. Each division can be done
with at most $O(n^2)$ operations hence the total number of
operations is $O(n^3)$.

There are basically three different types of Boolean networks:
classical deterministic, classical probabilistic, and quantum. They
correspond to, respectively, deterministic, randomised, and quantum
algorithms.

Classical deterministic networks are based on logical connectives
such as \textsc{and}, \textsc{or}, and \textsc{not} and are required
to always deliver
correct answers. If a
problem admits a deterministic uniform network family of polynomial
size, we say that the problem is in the class $P$~\cite{Pap94}.

Probabilistic networks have additional ``coin flip" gates which do
not have any inputs and emit one uniformly-distributed random
bit when executed during a computation. Despite the fact that
probabilistic networks may generate erroneous answers they may be
more powerful than deterministic ones. A good example is
primality testing -- given an $n$-bit number $x$ decide whether
or not $x$ is prime. The smallest known uniform deterministic network
family that solves this problem is of size $O(n^{d\log\log n})$,
which is not polynomially bounded. However, there is a
probabilistic algorithm, due to Solovay and Strassen~\cite{SS77}, that can
solve the same problem with a uniform probabilistic network family
of size $O(n^3\log (1/\epsilon))$, where $\epsilon$ is the
probability of error. \emph{N.B.} $\epsilon$ does not depend on $n$ and we
can choose it as small as we wish and still get an efficient
algorithm.

The $\log (1/\epsilon)$ part can be explained as follows. Imagine a
probabilistic network that solves a decision problem~\footnote{A
decision problem is a problem that admits only two answers: YES or
NO} and that errs with probability smaller than
$\frac{1}{2}+\delta$ for fixed $\delta>0$. If you run $r$ of these
networks in parallel (so that the size of the overall network is
increased by factor $r$) and then use the majority voting for the
final YES or NO answer your overall probability of error will
bounded by $\epsilon=\exp(-\delta^2 r)$. (This follows directly
from the Chernoff bound- see for instance, \cite{MR95}). Hence $r$
is of the order $\log (1/\epsilon)$. If a problem admits such a
family of networks then we say the problem is in the class $BPP$
(stands for ``bounded-error probabilistic
polynomial")~\cite{Pap94}.

Last but not least we have quantum algorithms, or families of
quantum networks, which are more powerful than their probabilistic
counterparts. The example here is the factoring problem -- given an
$n$-bit number $x$ find a list of prime factors of $x$. The
smallest known uniform probabilistic network family which solves
the problem is of size $O(2^{^{d\sqrt{n\log n}}})$. One reason why
quantum computation is such a fashionable field today is the
discovery, by Peter Shor, of a uniform family of quantum networks of
$O(n^2\log\log n\log(1/\epsilon))$ in
size, that solve the factoring
problem~\cite{Sho94}. If a problem admits a uniform quantum network
family of polynomial size that for any input gives the right answer
with probability larger than $\frac{1}{2}+\delta$ for fixed
$\delta>0$ then we say the problem is in the class $BQP$ (stands
for ``bounded-error quantum probabilistic polynomial").

We have
\begin{equation}
P\subseteq BPP \subseteq BQP
\end{equation}

Quantum networks are potentially more powerful because of
multiparticle quantum interference, an inherently quantum phenomenon
which makes
the quantum theory radically different from any classical
statistical theory.

Richard Feynman~\cite{Fey82} was the first to anticipate
the unusual power of quantum computers.
He observed that it appears to be impossible to simulate a general
quantum evolution on a classical probabilistic computer in an {\em
efficient} way \emph{i.e.} any classical simulation of quantum evolution
appears to involve an exponential slowdown in time as compared to
the natural evolution since the amount of information required to
describe the evolving quantum state in classical terms generally
grows exponentially in time.  However, instead of viewing this
fact as an obstacle, Feynman regarded it as an opportunity.
Let us then follow his lead and try to construct a computing
device using inherently quantum mechanical effects.

\section{From interferometers to computers}
A single particle interference in the Mach-Zehnder interferometer
works as follows.
A particle, in this case a photon, impinges on a beam-splitter
(BS1), and, with some probability amplitudes, propagates via two
different paths to another beam-splitter (BS2) which directs the
particle to one of the two detectors. Along each path between the
two beam-splitters, is a phase shifter (PS).

\setlength{\unitlength}{1mm}
\begin{picture}(80,60)(0,18)
\put(0,30){\line(1,0){30}}
\put(30,27){\framebox(5,6){$\phi_0$}}
\put(30,35){PS}
\put(35,30){\line(1,0){20}}
\put(10,20){\line(0,1){35}}
\put(55,30){\line(0,1){35}}
\put(10,55){\line(1,0){20}}
\put(30,52){\framebox(5,6){$\phi_1$}}
\put(30,60){PS}
\put(35,55){\line(1,0){30}}
\put(65,52){\line(0,1){6}}
\put(65,55){\oval(6,6)[r]}
\put(52,65){\line(1,0){6}}
\put(55,65){\oval(6,6)[t]}
\put(40,70){\makebox(30,6){$P_0=\cos^2\frac{\phi_0-\phi_1}{2}$}}
\put(69,52){\makebox(30,6){$P_1=\sin^2\frac{\phi_0-\phi_1}{2}$}}
{\thicklines
\put(8,28){\line(1,1){4}}
\put(53,53){\line(1,1){4}}
\put(11,26){BS1}
\put(56,51){BS2}
\put(8,53){\line(1,1){4}}
\put(53,28){\line(1,1){4}}}
\put(0,31){\makebox(5,5){$|0\rangle$}}
\put(4,20){\makebox(5,5){$|1\rangle$}}
\put(50,30){\makebox(5,5){$|0\rangle$}}
\put(10,50){\makebox(5,5){$|1\rangle$}}
\end{picture}

If the lower path is
labeled as state $\left | \, 0\right\rangle$ and the upper one as
state $\left |\, 1 \right \rangle$ then the particle, initially in
path $\left |\, 0 \right \rangle$, undergoes the following sequence
of transformations

\begin{eqnarray} \label{eqinterfere}
\left| \,0\right\rangle &\stackrel{\mbox{\tiny BS1}}{\mapsto }&\frac{
1}{\sqrt{2}}\left( \left| \,0\right\rangle +\left| \,1\right\rangle \right)
\stackrel{\mbox{\tiny
PS}}{\mapsto }\frac{1}{\sqrt{2}}(e^{i\phi _{0}}\left|
\,0\right\rangle +e^{i\phi _{1}}\left| \,1\right\rangle )\\
&=&e^{i\frac{\phi
_{0}+\phi _{1}}{2}}\frac{1}{\sqrt{2}}(e^{i\frac{\phi _{0}-\phi _{1}}{2}
}\left| \,0\right\rangle +e^{i\frac{-\phi _{0}+\phi _{1}}{2}}\left|
\,1\right\rangle )  \nonumber \\
&\stackrel{\mbox{\tiny BS2}}{\mapsto }&e^{i\frac{\phi
_{1}+\phi _{2} }{2}}(\cos \mbox{$\textstyle \frac{1}{2}$}(\phi
_{0}-\phi _{1})\left|
\,0\right\rangle +i\sin \mbox{$\textstyle \frac{1}{2}$}(\phi _{0}-\phi
_{1})\left| \,1\right\rangle ),
\end{eqnarray}
where $\phi _{0}$ and $\phi _{1}$ are the settings of the two phase shifters
and the action of the beam-splitters is defined as
\begin{equation}
\left| \,0\right\rangle {\mapsto }\textstyle{\frac{1}{\sqrt{2}}}
(\left| \,0\right\rangle +\left| \,1\right\rangle ),\quad
\left| \,1\right\rangle {\mapsto }\textstyle{\frac{1}{\sqrt{2}}}
(\left| \,0\right\rangle -\left| \,1\right\rangle ).  \label{tran}
\end{equation}
(We have ignored the phase shift in the reflected beam.) The global
phase shift $e^{i\frac{\phi _{0}+\phi _{0}}{2}}$ is irrelevant as
the interference pattern depends on the difference between the
phase shifts in different arms of the interferometer. The phase
shifters in the two paths can be tuned to effect any prescribed
relative phase shift $\phi =\phi _{0}-\phi _{1}$ and to direct the
particle with probabilities
\begin{eqnarray}
P_{0} &=&\cos ^{2}\left( \frac{\phi }{2}\right) =\frac{1}{2}\left( 1+\cos
\phi \right) \\
P_{1} &=&\sin ^{2}\left( \frac{\phi }{2}\right) =\frac{1}{2}\left( 1-\cos
\phi \right)
\end{eqnarray}
respectively to detectors ``0'' and ``1''.

The roles of the three key ingredients in this experiment are
clear. The first beam splitter prepares a superposition of possible
paths, the phase shifters modify quantum phases in different paths
and the second beam-splitter combines all the paths together
erasing all information about which path was actually taken by the
particle between the two beam-splitters. This erasure is very
important as we shall see in a moment.

Needless to say, single particle interference experiments are not
restricted to photons. One can go for a different ``hardware'' and
repeat the experiment with electrons, neutrons, atoms or even
molecules. When it comes to atoms and molecules both external and
internal degrees of freedom can be used.

Although single particle interference experiments are worth
discussing in their own right, here we are only interested in their
generic features simply because they are all ``isomorphic'' and
once you know and understand one of them you, at least for our
purposes, understand them all (modulo experimental details, of
course). Let us now describe any single particle interference
experiment in more general terms. It is very convenient to view
this experiment in a diagramatic way as a \emph{quantum network}
with three quantum logic gates~\cite{CEMM98}. The beam-splitters
will be now called the Hadamard gates and the phase shifters the
phase shift gates. In particular any single particle quantum
interference can be represented by the following simple network,
\setlength{\unitlength}{1mm}
\begin{center}
\begin{picture}(60,20)
\put(0,10){\line(1,0){7}}
\put(7,7){\framebox(6,6){$H$}}
\put(52,7){\framebox(6,6){$H$}}
\put(58,10){\line(1,0){10}}
\put(13,10){\line(1,0){39}}
\put(22,12){\makebox(21,6){$\phi=\phi_0-\phi_1$}}
\put(33,10){\circle*{3}}
\end{picture}
\end{center}

In order to make a connection with a quantum function evaluation
let us now describe an alternative construction which simulates
the action of the phase shift gate. This construction
introduces a phase factor $\phi$ using a controlled-$U$ gate.
The phase shift $\phi$ is ``computed'' with the help of an
auxiliary qubit in a prescribed state $\left| \,u\right\rangle $
such that $U\left|\,u\right\rangle =e^{i\phi
}\left|\,u\right\rangle$.

\setlength{\unitlength}{0.023 in}
\begin{center}
\begin{picture}(150,50)

\put(25,10){\makebox(0,0){$\ket{u}$}}
\put(25,40){\makebox(0,0){$\ket{0}$}}

\put(30,40){\line(1,0){10}}
\put(50,40){\line(1,0){40}}
\put(100,40){\line(1,0){10}}
\put(30,10){\line(1,0){35}}
\put(75,10){\line(1,0){35}}

\put(40,35){\framebox(10,10){$H$}}
\put(90,35){\framebox(10,10){$H$}}
\put(70,40){\circle*{4}}
\put(70,40){\line(0,-1){25}}
\put(65,5){\framebox(10,10){$U$}}

\put(115,10){\makebox(0,0){$\ket{u}$}}
\put(130,40){\makebox(0,0){\mbox{Measurement}}}
\end{picture}
\end{center}
In our example, shown above, we obtain the following sequence of
transformations on the two qubits
\begin{eqnarray}
\left | \, 0 \right \rangle\left | \, u \right \rangle \stackrel{H}{
\mapsto} \textstyle{\frac{1}{\sqrt 2}}(\left | \, 0 \right \rangle +
\left | \, 1 \right \rangle)\left | \, u \right \rangle & \stackrel{c-U}{
\mapsto} & \textstyle{\frac{1}{\sqrt 2}}(\left | \, 0 \right \rangle
+ e^{i\phi}\left | \, 1 \right \rangle) \left | \, u \right \rangle
\nonumber \\
& \stackrel{H}{\mapsto}& (\cos\textstyle{\frac{\phi }{2}}\left | \,
0 \right \rangle + i \sin\textstyle{\frac{\phi }{2}}\left | \, 1 \right
\rangle) \left | \, u \right \rangle.  \label{sequ}
\end{eqnarray}

We note that the state of the auxiliary qubit $\left| \,u\right\rangle $,
being an eigenstate of $U$, is not altered along this network, but its
eigenvalue $e^{i\phi }$ is ``kicked back'' in front of the $\left|
\,1\right\rangle $ component in the first qubit. The sequence (\ref{sequ})
is the exact simulation of the Mach-Zehnder interferometer and, as we shall
see later on, the kernel of quantum algorithms.

Some of the controlled-$U$ operations are special - they represent
quantum function evaluations! Indeed, a unitary evolution which
computes $f:\, \{0,1\}^n\mapsto \{0,1\}^m$,
\begin{equation}
\ket{x}\ket{y}\mapsto\ket{x}\ket{(y+f(x))\bmod 2^m},
\end{equation}
is of the controlled-$U$ type. The unitary transformation of the
second register, specified by
\begin{equation}
\ket{y}\mapsto\ket{(y+f(x))\bmod 2^m},
\end{equation}
depends on $x$ -- the state of the first register. If the initial
state of the second register is set to
\begin{equation}
\left| \,u\right\rangle = \frac{1}{2^{m/2}} \sum_{y=0}^{2^{m}-1}\exp \left( -\frac{2\pi i}{2^{m}
}y\right) |y\rangle,
\end{equation}
by applying the QFT to the state $\ket{111...1}$, then the function
evaluation generates
\begin{eqnarray}
|x\rangle \left| \,u\right\rangle &=& \frac{1}{2^{m/2}} |x\rangle \sum_{y=0}^{2^{m}-1}\exp
\left( -\frac{2\pi i}{2^{m}}y\right) |y\rangle \\
&\mapsto& \frac{1}{2^{m/2}} |x\rangle
\sum_{y=0}^{2^{m}-1}\exp \left( -\frac{2\pi i}{2^{m}}y\right) |f(x)+y\rangle
\\
&=& \frac{e^{ \frac{2\pi i}{2^{m}}f(x) }}{2^{m/2}} |x\rangle
\sum_{y=0}^{2^{m}-1}\exp \left( -\frac{2\pi i}{2^{m}}(f(x)+y)\right)
|f(x)+y\rangle \\
&=& \frac{e^{ \frac{2\pi i}{2^{m}}f(x) }}{2^{m/2}} |x\rangle
\sum_{y=0}^{2^{m}-1}\exp \left( -\frac{2\pi i}{2^{m}}y\right) |y\rangle \\
&=& e^{\frac{2\pi i}{2^{m}}f(x)} |x\rangle \left|
\,u\right\rangle ,
\end{eqnarray}
where we have relabelled the summation index in the sum containing
$2^{m}$ terms
\begin{equation}
\sum_{y=0}^{2^{m}-1}\exp \left( -\frac{2\pi i}{2^{m}}(f(x)+y)\right)
|f(x)+y\rangle =\sum_{y=0}^{2^{m}-1}\exp \left( -\frac{2\pi
i}{2^{m}} y\right) |y\rangle .
\end{equation}
Again, the function evaluation effectively introduces the phase
factors in front of the $|x\rangle $ terms in the first register.
\begin{equation}
|x\rangle \left| \,u\right\rangle \mapsto \exp \left(
\frac{2\pi i}{2^{m} }f(x)\right) |x\rangle \left| \,u\right\rangle
\end{equation}
Please notice that the resolution in $\phi (x)=\frac{2\pi
}{2^{m}}f(x)$ is determined by the size $m$ of the second register.
For $m=1$ we obtain $\phi (x)=\pi f(x)$, \emph{i.e.} the phase factors are
$(-1)^{f(x)}$. Let us see how this approach explains the internal
working of quantum algorithms.

\section{The first quantum algorithms}
The first quantum algorithms showed advantages of quantum
computation without referring to computational complexity measured
by the scaling properties of network sizes. The computational power
of quantum interference was discovered by counting how many times
certain Boolean functions have to be evaluated in order to find the
answer to a given problem. Imagine a ``black box" (also called an
\emph{oracle}) computing a Boolean function and a scenario in which one
wants to learn about a given property of the Boolean function but
has to pay for each use of the ``black box" (often referred to as a
\emph{query}). The objective is to minimise number of queries.

Consider, for example, a ``black box" computing a Boolean function
$f:\, \{0,1\}\mapsto\{0,1\}$. There are exactly four such
functions: two constant functions ($f(0)=f(1)=0$ and $f(0)=f(1)=1$)
and two ``balanced'' functions ($ f(0)=0,f(1)=1$ and
$f(0)=1,f(1)=0$). The task is to deduce, by queries to the ``black
box", whether $f$ is constant or balanced (in other words, whether
$f(0)$ and $f(1)$ are the same or different).

Classical intuition tells us that we have to evaluate both $f(0)$
and $f(1)$, which involves evaluating $f$ twice (two queries). We
shall see that this is not so in the setting of quantum
information, where we can solve this problem with a single function
evaluation (one query), by employing an algorithm that has the same
mathematical structure as the Mach-Zehnder interferometer. The
quantum algorithm that accomplishes this is best represented as the
quantum network shown below, where the middle operation is the
``black box" representing the function evaluation~\cite{CEMM98}.
\setlength{\unitlength}{0.025 in}
\begin{center}
\begin{picture}(150,50)

\put(15,10){\makebox(0,0){$\ket{0}-\ket{1}$}}
\put(22,40){\makebox(0,0){$\ket{0}$}}

\put(30,40){\line(1,0){10}}
\put(50,40){\line(1,0){40}}
\put(100,40){\line(1,0){10}}
\put(30,10){\line(1,0){35}}
\put(75,10){\line(1,0){35}}

\put(40,35){\framebox(10,10){$H$}}
\put(90,35){\framebox(10,10){$H$}}
\put(70,40){\circle*{4}}
\put(70,40){\line(0,-1){25}}
\put(65,5){\framebox(10,10){$f$}}

\put(124,10){\makebox(0,0){$\ket{0}-\ket{1}$}}
\put(129,40){\makebox(0,0){\mbox{Measurement}}}
\end{picture}
\end{center}
The initial state of the qubits in the quantum network is $\left|
\,0\right\rangle (\left| \,0\right\rangle -\left| \,1\right\rangle
)$ (apart from a normalization factor, which will be omitted in the
following). After the first Hadamard transform, the state of the
two qubits has the form $(\left| \,0\right\rangle +\left|
\,1\right\rangle )(\left| \,0\right\rangle-\left| \,1\right\rangle
)$. To determine the effect of the function evaluation on this
state, first recall that, for each $x\in \{0,1\}$,

\begin{equation}
\left| \,x\right\rangle (\left| \,0\right\rangle -\left| \,1\right\rangle )
\stackrel{f}{\mapsto }\,(-1)^{f(x)}\left| \,x\right\rangle (\left|
\,0\right\rangle -\left| \,1\right\rangle ).
\end{equation}
Therefore, the state after the function evaluation is
\begin{equation}
\lbrack (-1)^{f(0)}\left| \,0\right\rangle +(-1)^{f(1)}\left|
\,1\right\rangle ](\left| \,0\right\rangle -\left| \,1\right\rangle )\;.
\label{st-d}
\end{equation}
That is, for each $x$, the $\left| \,x\right\rangle $ term acquires a phase
factor of $(-1)^{f(x)}$, which corresponds to the eigenvalue of the state of
the auxiliary qubit under the action of the operator that sends $\left|
\,y\right\rangle $ to $\left| \,y+f(x)\right\rangle $. The second qubit is
of no interest to us any more but the state of the first qubit
\begin{equation}
(-1)^{f(0)}\left| \,0\right\rangle +(-1)^{f(1)}\left| \,1\right\rangle
\end{equation}
is equal either to
\begin{equation}
\pm \left( \left| \,0\right\rangle +\left| \,1\right\rangle \right) ,
\end{equation}
when $f(0)=f(1),$ or
\begin{equation}
\pm \left( \left| \,0\right\rangle -\left| \,1\right\rangle \right) ,
\end{equation}
when $f(0)\neq f(1).$ Hence, after applying the second Hadamard
gate the state of the first qubit becomes $\left|
\,0\right\rangle $ if the function $f$ is constant and $\left|
\,1\right\rangle $ if the function is balanced! A bit-value
measurement on this qubit distinguishes these cases with certainty.

This example~\cite{CEMM98} is an improved version of the first
quantum algorithm proposed by Deutsch~\cite{Deu85} (The original
Deutsch algorithm provides the correct answer with probability
50\%.) Deutsch's result laid the foundation for the new field of
quantum computation, and was followed by several other quantum
algorithms.

Deutsch's original problem was subsequently generalised to cover
``black boxes" computing Boolean functions $f
:\{0,1\}^n\mapsto\{0,1\}$. Assume that, for one of these
functions, it is ``promised'' that it is either constant or
balanced (\emph{i.e.} has an equal number of 0's outputs as 1's), and the
goal is to determine which of the two properties the function
actually has. How many queries to $f$ are required to do this? Any
classical algorithm for this problem would, in the worst-case,
require $2^{n-1}+1$ queries before determining the answer with
certainty. There is a quantum algorithm that solves this problem
with a single evaluation of $f$.

The algorithm is illustrated by a simple extension of the network
which solves Deutsch's problem.
\begin{center}
\setlength{\unitlength}{0.027 in}
\begin{picture}(110,60)

\put(5,54){$\ket{0}$}
\put(20,55){\line(1,0){10}}
\put(30,50){\framebox(10,10){$H$}}
\put(40,55){\line(1,0){20}}
\put(60,50){\framebox(10,10){$H$}}
\put(70,55){\line(1,0){10}}
\put(50,55){\circle*{3}}
\put(85,54){Measurement}

\put(5,39){$\ket{0}$}
\put(20,40){\line(1,0){10}}
\put(30,35){\framebox(10,10){$H$}}
\put(40,40){\line(1,0){20}}
\put(60,35){\framebox(10,10){$H$}}
\put(70,40){\line(1,0){10}}
\put(85,39){Measurement}

\put(5,24){$\ket{0}$}
\put(20,25){\line(1,0){10}}
\put(30,20){\framebox(10,10){$H$}}
\put(40,25){\line(1,0){20}}
\put(60,20){\framebox(10,10){$H$}}
\put(70,25){\line(1,0){10}}
\put(50,25){\circle*{3}}
\put(85,24){Measurement}

{\linethickness{0.081 in}
\put(50,55){\line(0,-1){30}}
}

\put(0,4){$\ket{0}-\ket{1}$}
\put(20,5){\line(1,0){25}}
\put(45,0){\framebox(10,10){$f$}}
\put(55,5){\line(1,0){25}}
\put(85,4){$\ket{0}-\ket{1}$}

\put(50,25){\line(0,-1){15}}

\end{picture}
\end{center}
The control register, now composed out of $n$ qubits ($n=3$ in the
diagram above), is initially in state $\ket{00\cdots 0}$ and an
auxiliary qubit in the second register starts and remains in the
state $\ket{0}-\ket{1}$.

Stepping through the execution of the network, the state after the
first $n$-qubit Hadamard transform is applied is
\begin{equation}
  \sum_{x}\ket{x}(\ket{0}-\ket{1})\;,
\end{equation}
which, after the function evaluation, is
\begin{equation}
  \sum_{x}(-1)^{f(x)}\ket{x}(\ket{0}-\ket{1}).
\end{equation}
Finally, after the last Hadamard transform, the state is
\begin{equation}
  \sum_{x,y}(-1)^{f(x)+(x \cdot y)}\ket{y}
  (\ket{0}-\ket{1}).
\end{equation}

Note that the amplitude of $\ket{00\cdots 0}$ is $\sum_{x}
\frac{(-1)^{f(x)}}{2^n}$ which is $(-1)^{f(0)}$ when $f$ is
constant and $0$ when $f$ is balanced. Therefore, by measuring the
first $n$ qubits, it can be determined with certainty whether $f$
is constant or balanced. The algorithm follows the same pattern as
Deutsch's algorithm: the Hadamard transform, a function evaluation,
the Hadamard transform (the H-f-H sequence). We recognize it as a
generic interference pattern.

\section{Quantum search}
The generic H-f-H sequence may be repeated several times. This can
be illustrated, for example, with Grover's data base search
algorithm~\cite{Gro96}. Suppose we are given, as an oracle, a
Boolean function $f_k$ which maps $\{0,1\}^n$ to $\{0,1\}$ such
that $f_k(x)=\delta_{xk}$ for some $k$. Our task is to find $k$.
Thus in a set of numbers from $0$ to $2^{n}-1$ one element has been
``tagged'' and by evaluating $f_k$ we have to find which one. In
order to find $k$ with probability of $50\%$ any classical
algorithm, be it deterministic or randomised, will need to evaluate
$f_k$ a minimum of $2^{n-1}$ times. In contrast, a quantum
algorithm needs only $O(2^{n/2})$ evaluations.

Unlike the algorithms studied so far, Grover's algorithm
consists of \emph{repeated} applications of the \emph{same}
unitary transformation many ($O(2^{n/2})$) times.  The initial
state is chosen to be the one that has equal overlap with each of
the computational basis states: $\ket{S} = 2^{-n/2}\sum_{i=0}^{2^n} \ket{i}$.
The operation applied at each individual iteration, referred to as
the Grover iterate,
can be best represented by the following network:
\setlength{\unitlength}{0.025 in}
\begin{center}
\begin{picture}(100,50)

\put(15,40){\line(1,0){25}}
\put(50,40){\line(1,0){20}}
\put(80,40){\line(1,0){10}}

\put(40,35){\framebox(10,10){$H$}}
\put(60,40){\circle*{4}}
\put(70,35){\framebox(10,10){$H$}}
\put(30,40){\circle*{4}}

\put(15,10){\line(1,0){10}}
\put(60,40){\line(0,-1){25}}
\put(55,5){\framebox(10,10){$f_0$}}
\put(35,10){\line(1,0){20}}
\put(65,10){\line(1,0){25}}
\put(30,40){\line(0,-1){25}}
\put(25,5){\framebox(10,10){$f_k$}}

\put(0,40){\makebox(0,0){$\ket{\psi}$}}
\put(0,10){\makebox(0,0){$\ket{0}-\ket{1}$}}

\end{picture}
\end{center}
The components of the network are by now familiar:
Hadamard transforms ($H$) and
controlled-$f$ gates.  It is important to notice that in drawing
the network we have used a shorthand notation: the first register
(with the $\ket{\psi}$ input) actually consists of $n$ qubits.  The
Hadamard transform is applied to each of those qubits and the
controlled-$f$ gates act on all of them simultaneously.
Also, the input to the second register is
always $\ket{0}-\ket{1}$ but the input to the first register,
denoted $\ket{\psi}$ changes from iteration from iteration, as the
calculation proceeds.
As usual, the second register will be ignored since
it remains constant throughout the computation.

To begin, consider only the
controlled-$f_k$ gate.
This is just the phase-kickback construction that was
introduced in Section 4 but for the specific function $f_k$.
In particular, the transformation does nothing to any basis
elements except for
$\ket{k}$, which goes to $-\ket{k}$.  Geometrically, this is
simply a reflection in the hyperplane perpendicular to $\ket{k}$
so let us call it $R_k$.

Similarly, with respect to the first register only, the
controlled-$f_0$ operation sends $\ket{0}$ to $-\ket{0}$
and fixes all other basis elements, so it can be written $R_0$.
Now consider the sequence of operations
$H R_0 H$.  Since $H^2 = I$, we can rewrite the triple as
$H R_0 H^{-1}$ which is simply $R_0$ performed in a different
basis.  More specifically, it is reflection about the hyperplane
perpendicular to
\begin{equation}
H \ket{0} = \frac{1}{2^{n/2}} \sum_{x=0}^{2^n-1} \ket{x} = \ket{S}
\end{equation}
so we will simply write the triple as $R_S$.

We can therefore rewrite the Grover iterate in the simple form
$G = R_S R_k$.
Now, since each reflection is an orthogonal transformation with
negative determinant, their composition must be an orthogonal
transformation with unit determinant, in other words, a rotation.
The question, of course, is which rotation.  To find the answer
it suffices to consider rotations in the plane spanned by
$\ket{k}$ and $\ket{S}$ since all other vectors are fixed by the
Grover iterate.  The generic geometrical situation is then illustrated in the
following diagram.
\begin{center}
\includegraphics[height=0.35\hsize]{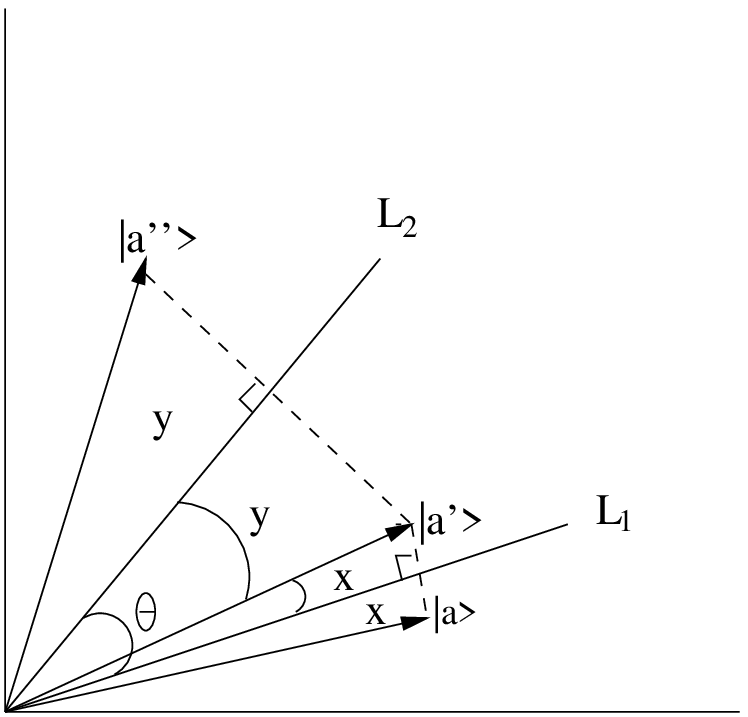}
\end{center}
If the vector $\ket{a}$ is reflected through the line $L_1$ to produce
the vector $\ket{a'}$ and then reflected a second time through
line $L_2$ to produce the vector $\ket{a''}$, then the net effect is
a rotation by the total subtended angle between $\ket{a}$ and $\ket{a''}$,
which is $2 x + 2 y = 2 ( x + y ) = 2 \theta$.

Therefore, writing $\ket{k^\perp}$ and $\ket{S^\perp}$ for plane vectors
perpendicular to $\ket{k}$ and $\ket{S}$ respectively, the
Grover iterate performs a rotation of twice the angle from
$\ket{k^\perp}$ to $\ket{S^\perp}$.  Setting,
$\sin \phi = \frac{1}{2^{n/2}}$, this is easily seen to be a rotation
by
\begin{equation}
2( 3 \frac{\pi}{2} - \phi) = \pi - 2 \phi \bmod 2 \pi.
\end{equation}
Thus, up to phases, the Grover iterate rotates the state
vector by an angle $2 \phi$ towards the desired solution
$\ket{k}$.
Normally, the initial state for the first register is chosen
to be $\ket{S}$.  Since this initial state $\ket{S}$ is already
at an angle $\phi$ to $\ket{k}$, the iterate should be
repeated
$m$ times, where
\begin{equation}
(2 m + 1) \phi \approx \frac{\pi}{2},
\end{equation}
giving
\begin{equation}
m \approx \frac{\pi}{4 \phi} - \frac{1}{4}
\end{equation}
to get a probability of success bounded below by $\cos^2 (2\phi)$,
which goes to 1 as $n \mapsto \infty$.
For large $n$, $\frac{1}{2^{n/2}} = \sin \phi \approx \phi$,
so
\begin{equation}
m \approx \frac{\pi}{4} \frac{1}{2^{n/2}}.
\end{equation}

This is an astounding result: any search of an unstructured
database can be performed in time proportional to the square-root
of the number of entries in the database.  Subsequent work
extended the result to searches for multiple items~\cite{Boy96}, searches
of structured databases~\cite{Hog98}, and many other situations.
Also, Zalka~\cite{Zal99}, Boyer et. al~\cite{Boy96}
and others have demonstrated that Grover's
algorithm is optimal, in the sense that any other quantum algorithm
for searching an unstructured database must take time at least
$O(2^{n/2})$.

\section{Optimal phase estimation}
Query models of quantum computation provided a natural setting for
subsequent discoveries of ``real quantum algorithms". The most
notable example is Shor's quantum factoring algorithm~\cite{Sho94}
which evolved from the the order-finding problem, which was
originally formulated in the language of quantum queries. Following
our ``interferometric approach" we will describe this algorithm in
the terms of multiparticle quantum interferometry. We start with a
simple eigenvalue or phase estimation problem.

Suppose that $U$ is any unitary transformation on $m$ qubits and
$\left | \, u \right \rangle$ is an eigenvector of $U$ with
eigenvalue $e^{i\phi}$ and consider the following scenario. We do
not explicitly know $U$ or $\left | \, u\right \rangle$ or $e^{i
\phi}$, but instead we are given devices that perform controlled-$U$,
controlled-$U^{2^1}$, controlled-$U^{2^2}$ and so on until we reach
controlled-$U^{2^{n-1}}$. Also, assume that we are given a single
preparation of the state $\left | \, u\right\rangle$. Our goal is
to obtain an $n$-bit estimator of $\phi$. We start by constructing
the following network,

\medskip
\setlength{\unitlength}{0.01 in}
\begin{center}
\begin{picture}(305,180)(105,380)
\thicklines
\put(155,380){\framebox(30,60){$U^{2^{0}}$}}
\put(215,380){\framebox(30,60){$U^{2^{1}}$}}
\put(280,380){\framebox(30,60){$U^{2^{2}}$}}
\put(110,425){\line( 1, 0){ 45}}
\put(185,425){\line( 1, 0){ 30}}
\put(245,425){\line( 1, 0){ 35}}
\put(310,425){\line( 1, 0){ 75}}
\put(110,415){\line( 1, 0){ 45}}
\put(185,415){\line( 1, 0){ 30}}
\put(245,415){\line( 1, 0){ 35}}
\put(310,415){\line( 1, 0){ 75}}
\put(110,405){\line( 1, 0){ 45}}
\put(185,405){\line( 1, 0){ 30}}
\put(245,405){\line( 1, 0){ 35}}
\put(310,405){\line( 1, 0){ 75}}
\put(110,395){\line( 1, 0){ 45}}
\put(185,395){\line( 1, 0){ 30}}
\put(245,395){\line( 1, 0){ 35}}
\put(310,395){\line( 1, 0){ 75}}
\put(105,560){\line( 1, 0){280}}
\put(105,525){\line( 1, 0){280}}
\put(105,485){\line( 1, 0){280}}
\put(170,440){\line( 0, 1){ 45}}
\put(230,440){\line( 0, 1){ 85}}
\put(295,440){\line( 0, 1){120}}
\put(80,407){$\ket{u}$}
\put(397,407){$\ket{u}$}
\put(50,485){$\ket{0}+\ket{1}$}
\put(50,525){$\ket{0}+\ket{1}$}
\put(50,560){$\ket{0}+\ket{1}$}
\put(393,485){$\ket{0}+e^{i2^{0}\phi}\ket{1}$}
\put(393,525){$\ket{0}+e^{i2^{1}\phi}\ket{1}$}
\put(393,560){$\ket{0}+e^{i2^{2}\phi}\ket{1}$}
\put(170,485){\circle*{6}}
\put(230,525){\circle*{6}}
\put(295,560){\circle*{6}}
\end{picture}
\end{center}
\medskip

The second register of $m$ qubits is initially prepared in state
$\ket{u}$ and remains in this state after the computation, whereas
the first register of $n$ qubits evolves into the state,
\begin{equation}
(\left | \, 0 \right \rangle + e^{i 2^{n-1} \phi}\left | \, 1 \right
\rangle) (\left | \, 0 \right \rangle + e^{i 2^{n-2} \phi}\left | \, 1
\right \rangle) \cdots (\left | \, 0 \right \rangle + e^{i \phi}\left | \, 1
\right \rangle)=\sum_{y=0}^{2^n-1} e^{ 2\pi i \frac{\phi y}{2^n}} \ket{y}.
\label{qftphi}
\end{equation}

Consider the special case where $\phi = 2\pi x/2^{n}$ for
$x=\sum_{i=0}^{n-1}2^i x_i$, and recall the quantum Fourier transform
(QFT) introduced in Section 2. The state which gives the binary 
representation of $x$,
namely, $\left |\, x_{n-1}\cdots x_0 \right\rangle$ (and hence
$\phi$) can be obtained by applying the inverse of the QFT , that
is by running the network for the QFT in the backwards direction
(consult the diagram of the QFT). If $x$ is an $n$-bit number this
will produce the exact value $\phi$.

However, $\phi$ does not have to be a fraction of a power of two
(and may not even be a rational number). For such a $\phi$, it
turns out that applying the inverse of the QFT produces the best
$n$-bit approximation of $\phi$ with probability at least $4 /
\pi^2\approx 0.405 $.

To see why this is so, let us write $\phi = 2\pi (a/2^{n}+\delta)$,
where $a=(a_{n-1}\ldots a_{0})$ is the best $n$-bit estimate of
$\frac{\phi}{2\pi}$ and $0 < |\delta| \le 1/2^{n+1}$. Applying the
inverse QFT to the state in Eq.~(\ref{qftphi}) now yields the state
\begin{equation}
{1\over 2^n} \sum_{x=0}^{2^n-1} \sum_{y=0}^{2^n-1} e^{\frac{2\pi
i}{2^n} (a-x) y} e^{2\pi i \delta y}\ket{x}
\label{qftstate}
\end{equation}
and the coefficient in front of $\ket{x=a}$ in the above is the
geometric series
\begin{eqnarray}
  {1 \over 2^n} \sum_{y=0}^{2^n-1} (e^{2\pi i \delta})^y & = & {1
    \over 2^n} \left({1 - (e^{2\pi i \delta})^{2^n} \over 1 - e^{2\pi
      i \delta}}\right)\;.
\end{eqnarray}
Since $|\delta| \le {1 \over 2^{n+1}}$, it follows that
$2^n|\delta| \le 1/2$, and using the inequality $2 z \leq \sin \pi z \leq \pi z$ holding for any $z\in[0,1/2]$, we get $|1 - e^{2\pi i \delta 2^n}| = 2|\sin (\pi\delta 2^n)| \ge  4 |\delta| 2^n$.  Also, $|1 - e^{2 \pi i \delta}|=2|\sin \pi\delta| \le 2 \pi \delta$.
Therefore, the probability of observing $a_{n-1} \cdots a_0$ when measuring the state is
\begin{equation}
  \left|{1 \over 2^n} \left({1 - (e^{2\pi i \delta})^{2^n} \over 1 -
    e^{2\pi i \delta}}\right)\right|^2 \ge \left({1 \over 2^n}
  \left({4 \delta 2^n \over 2 \pi \delta}\right)\right)^2 = {4 \over
    \pi^2},
\end{equation}
which proves our assertion. In fact, the probability of obtaining
the best estimate can be made $1-\delta$ for any $0<\delta <1$, by
creating the state in Eq.(\ref{qftphi}) but with
$n+O(\log(1/\delta))$ qubits and rounding the answer off to the
nearest $n$ bits~\cite{CEMM98}.

\section{Periodicity and quantum factoring}
Amazingly, the application of optimal phase estimation
to a very particular unitary operator will allow us to
factor integers efficiently.  In fact, it will allow
us to solve a more general class of problems related
to the periodicity of certain integer functions.

Let $N$ be an $m$-bit integer, and let $a$ be an
integer smaller than $N$, and coprime to $N$. Define a unitary
operator $U_a$ acting on $m$ qubits such that for all $y < N$
\begin{equation}
\quad \ket{y} \mapsto U_a \ket{y} = \ket{ a y \bmod N}.
\end{equation}
This unitary operation can be called multiplication by $a$ modulo
$N$. Since $a$ is coprime to $N$, as discussed in Section 2,
there exists a least strictly
positive $r$ such that $a^r =1 \bmod N$.  This $r$ is called
the \emph{order} of $a$ modulo $N$. Equivalently, $r$ is
the period of the function $f(x)=a^x \bmod N$, \emph{i.e.} the
least $r > 0$ such that $f(x) = f(x+r)$ for all $x$. We
are after the optimal $n$-bit estimate of this period, given
some specified precision $n$.

Now let the vectors $\ket{u_k}$ ($k\in\{1,\ldots,r\}$) be defined by
\begin{equation}
\ket{ u_{k}}= r^{-1/2} \sum_{j=0}^{r-1} e^{-\frac{2\pi i k j}{r}}\ket{a^j \bmod N}.
\end{equation}
It is easy to check~\cite{Kit95} that for each
$k\in\{1,\ldots,r\}$, $\ket{u_k}$ is an eigenvector with
eigenvalue $e^{2 \pi i {\frac{k}{r}} }$ of the
modular multiplication
operator $U_a$ defined above.

It is important to observe that one can efficiently construct a
quantum network for controlled multiplication modulo some number
$N$. Moreover, for any $j$, it is possible to efficiently implement
a controlled-$U^{2^j}_a$ gate~\cite{VBE96,BCDP96}. Therefore, we
can apply the techniques for optimal phase estimation discussed in
Section 7. For any $k\in\{1,\ldots,r \}$, given the state
$\ket{u_k}$ we can obtain the best $n$-bit approximation to
$\frac{k}{r}$. This is tantamount to determining $r$ itself.
Unfortunately, there is a complication.

Our task is: given an $m$ bit long number $N$ and randomly chosen
$a<N$ coprime with $N$, find the order of $a$ modulo $N$. The
problem with the above method is that we are aware of no
straightforward efficient way to prepare any of the states $\left |
\, u_k\right
\rangle$. However, the state
\begin{equation}
\left | \, 1 \right \rangle = r^{-1/2} \sum_{k=1}^{r} \left | \, u_k \right
\rangle
\end{equation}
\emph{is} most definitely an easy state to prepare.

If we start with $\ket{1}$ in place of the eigenvector $\ket{u_k}$,
apply the phase estimation network and measure the first register
bit by bit we will obtain $n$ binary digits of $x$ such that, with
probability exceeding $4/\pi^2$, $\frac{x}{2^n}$ is the best
$n$-bit estimate of ${\frac{k}{r}}$ for a randomly chosen $k$ from
$\{1,\ldots,r\}$. The question is: given $x$ how to compute $r$?
Let us make few observations:
\renewcommand{\labelenumi}{$\bullet$}
\begin{enumerate}
\item \emph{$k/r$ is unique, given $x$.} \\
Value $x/2^n$, being the $n$-bit estimate, differs by at most $1/2^n$
from $k/r$. Hence, as long as $n>2m$, the $n$ bit estimate $x$
determines a unique value of ${\frac{k}{r}}$ since $r$ is an 
$m$-bit number. 
\item \emph{Candidate values for $k/r$ are all convergents to $x/2^m$.} \\
For any real number $\theta$, there is a unique sequence of
special rationals
$(\frac{p_n}{q_n})_{n\in{\bf N}}$ ($\gcd(p_n,q_n)=1$) called
the \emph{convergents} to $\theta$ that tend to $\theta$ as $n$ grows.
A theorem~\cite{HW79} states that if $p$ and $q$ are integers
with $\left|\theta-\frac{p}{q}\right| < \frac{1}{2q^2}$ then $p/q$ is a
convergent to $\theta$. Since we have
$\frac{1}{2^n} \leq \frac{1}{2 (2^m)^2} \leq \frac{1}{2 r^2}$, this implies
$\left|\frac{x}{2^n}-\frac{k}{r}\right| < \frac{1}{2 r^2}$ and $k/r$ is
a convergent to $x/2^n$.
\item \emph{Only one convergent is eligible.} \\
It is easy to show that there is at most one fraction $a/b$ 
satisfying both $b \leq r$ and
$\left|\frac{x}{2^n}-\frac{a}{b}\right| < \frac{1}{2 r^2}$.
\end{enumerate}

Convergents can be found efficiently using the well-known 
\emph{continued fraction} method~\cite{HW79}. 
Thus we employ continued fractions and our observations above
to find a fraction $a/b$ such that $b\leq 2^m$ and
$\left|\frac{x}{2^n}-\frac{a}{b}\right| < \frac{1}{2^n}$.  We get
the rational $k/r$, and $k=a, r=b$, provided $k$ and $r$ are
coprime. For randomly chosen $k$, this happens with probability
greater than or equal to $1/\ln r$~\cite{EJ96}.

Finally, we show how order-finding can be used to factor a
composite number $N$. Let $a$ be a randomly chosen positive integer
smaller than $N$ such that $\gcd(a,N)=1$. Then the order of $a$
modulo $N$ is defined, and we can find it efficiently using the
above algorithm. If $r$ is even, then we have:
\begin{eqnarray}
a^r &=& 1\bmod N\\
\Leftrightarrow \quad (a^{r/2})^2 -1^2 &=& 0\bmod N\\
\Leftrightarrow \quad (a^{r/2}-1)(a^{r/2}+1)&=&0\bmod N.
\end{eqnarray}

The product $(a^{r/2}-1)(a^{r/2}+1)$ must be some multiple of $N$,
so unless $a^{r/2}=\pm 1\bmod N$ at least one of terms must have a
nontrivial factor in common with $N$. By computing the 
greatest common divisor of this
term and $N$, one gets a non-trivial factor of $N$.

Furthermore, if $N$ is odd with prime factorisation
\begin{equation}
N=p_1^{\alpha_1}p_2^{\alpha_2}\cdots p_s^{\alpha_s},
\end{equation}
then it can be shown~\cite{EJ96} that if $a<N$ is chosen at random
such that $\gcd(a,N)=1$ then the probability that its order modulo
$N$ is even and that $a^{r/2}\neq \pm 1\bmod N$ is:
\begin{equation}
\Pr(r \mbox{ is even \textsc{and} }
a^{r/2}\neq \pm 1 \bmod N) \ge 1-\frac{1}{2^{s-1}}.
\label{prob}
\end{equation}
Thus, combining our estimates of success at each step,
with probability greater than or equal to
\begin{equation}
\frac{4}{\pi^2} \frac{1}{\ln r} \left( 1 - \frac{1}{2^{s-1}} \right)
\ge \frac{2}{\pi^2} \frac{1}{\ln N}
\end{equation}
we find a factor of $N$~\footnote{\emph{N.B.} by Eq.(\ref{prob}), the
method fails if $N$ is a prime power, $N=p^\alpha$, but prime
powers can be efficiently recognised and factored by classical
means.}. (Here we have used that $N$ is composite and $r < N$.) If
$N$ is $\log N= n$ bits long then by repeating the whole process
$O(n)$ times, or by a running $O(n)$ computations in parallel by a
suitable extension of a quantum factoring network, we can then
guarantee that we will find a factor of $N$ with a fixed
probability greater than $\frac{1}{2}$. This, and the fact that the
quantum network family for controlled multiplication modulo some
number is uniform and of size $O(n^2)$, tells us that factoring is
in the complexity class $BQP$.

But why should anybody care about efficient factorisation?

\section{Cryptography}
Human desire to communicate secretly is at least as old as writing
itself and goes back to the beginnings of our civilisation. Methods
of secret communication were developed by many ancient societies,
including those of Mesopotamia, Egypt, India, and China, but
details regarding the origins of cryptology\footnote{The science of
secure communication is called cryptology from Greek \emph{kryptos}
hidden and \emph{logos} word. Cryptology embodies cryptography, the
art of code-making, and cryptanalysis, the art of code-breaking.}
remain unknown~\cite{Kah67}.

Originally the security of a cryptosystem or a cipher depended on the secrecy of
the entire encrypting and decrypting procedures; however, today we
use ciphers for which the algorithm for encrypting and decrypting
could be revealed to anybody without compromising their security. In such ciphers a set of specific
parameters, called a \textit{key}, is supplied together with the
plaintext as an input to the encrypting algorithm, and together
with the cryptogram as an input to the decrypting algorithm~\cite{Sti95}. This
can be written as
\begin{equation}
\hat E_{k}(P) = C, \; \mathrm{and\; conversely,}\; \hat D_{k}(C)=P,
\end{equation}
where $P$ stands for plaintext, $C$ for cryptotext or cryptogram,
$k$ for cryptographic key, and $\hat E$ and $\hat D$ denote an
encryption and a decryption operation respectively.

The encrypting and decrypting algorithms are publicly known; the
security of the cryptosystem depends entirely on the secrecy of the
key, and this key must consist of a \textit{randomly chosen},
sufficiently long string of bits. Probably the best way to explain
this procedure is to have a quick look at the Vernam cipher, also
known as the one-time pad~\cite{Ver26}.

If we choose a very simple digital alphabet in which we use only
capital letters and some punctuation marks such as

\setlength{\tabcolsep}{1.5 mm}
\begin{center}
\begin{tabular}
[c]{|ccccccccccccccc|}\hline
A & B & C & D & E & ... & \  & ... & X & Y & Z & \  & ? & , & .\\
00 & 01 & 02 & 03 & 04 & ... & \  & ... & 23 & 24 & 25 & 26 & 27 & 28 &
29\\\hline
\end{tabular}
\end{center}

we can illustrate the secret-key encrypting procedure by the
following simple example (we refer to the dietary requirements of
007):

{\scriptsize
\begin{center}
\begin{tabular}
[t]{cccccccccccccccccc}
S & H & A & K & E & N &  & N & O & T &  & S & T & I & R & R & E & D\\
18 & 07 & 00 & 10 & 04 & 13 & 26 & 13 & 14 & 19 & 26 & 18 & 19 & 08 & 17 &
17 & 04 & 03\\
15 & 04 & 28 & 13 & 14 & 06 & 21 & 11 & 23 & 18 & 09 & 11 & 14 & 01 & 19 &
05 & 22 & 07\\
03 & 11 & 28 & 23 & 18 & 19 & 17 & 24 & 07 & 07 & 05 & 29 & 03 & 09 & 06 &
22 & 26 & 10
\end{tabular}
\end{center}
}

In order to obtain the cryptogram (sequence of digits in the bottom
row) we add the plaintext numbers (the top row of digits) to the
key numbers (the middle row), which are randomly selected from
between 0 and 29, and take the remainder after division of the sum
by 30, that is we perform addition modulo 30. For example, the
first letter of the message ``S'' becomes a number ``18''in the
plaintext, then we add $18+15=33;\ 33=1\times30+3$, therefore we
get 03 in the cryptogram. The encryption and decryption can be
written as $P_i+k_i \pmod{30}=C_i$ and
$C_i-k_i \pmod{30}=P_i$ respectively for the symbol at position $i$.

The cipher was invented in 1917 by the American AT\&T engineer
Gilbert Vernam. It was later shown, by Claude
Shannon~\cite{Sha49}, that as long as the key is truly random,
has the same length as the message, and is never reused then the
one-time pad is perfectly secure. So, if we have a truly
unbreakable system, what is wrong with classical cryptography?

There is a snag. It is called \emph{key distribution}. Once the key
is established, subsequent communication involves sending
cryptograms over a channel, even one which is vulnerable to total
passive eavesdropping (\emph{e.g.} public announcement in mass-media).
This stage is indeed secure. However in order to establish the key,
two users, who share no secret information initially, must at a
certain stage of communication use a reliable and a very secure
channel. Since the interception is a set of measurements performed
by an eavesdropper on this channel, however difficult this might be
from a technological point of view, \textit{in principle} any
{classical} key distribution can always be passively monitored,
without the legitimate users being aware that any eavesdropping has
taken place.

In the late 1970s Whitfield Diffie and Martin Hellman~\cite{DH76b}
proposed an interesting solution to the key distribution problem.
It involved two keys, one public key $\pi$ for encryption and one
private key $\kappa$ for decryption:
\begin{equation}
\hat E_\pi(P) = C, \; \mathrm{and\; }\; \hat D_\kappa(C)=P.
\end{equation}
In these systems users do not need to
share any private key before they start sending messages to each other.
Every user has his own two keys; the public key is publicly
announced and the private key is kept secret. Several public-key
cryptosystems have been proposed since 1976; here we concentrate
our attention on the most popular one namely the RSA~\cite{RSA78}. In fact the techniques were first discovered at CESG in the
early 1970s by James Ellis, who called them ``Non-Secret
Encryption''~\cite{Ell70}. In 1973, building on Ellis' idea, C. Cocks designed
what we now call RSA~\cite{Coc73}, and in 1974 M. Williamson proposed what is
essentially known today as the Diffie-Hellman key exchange
protocol.

Suppose that Alice wants to send an RSA encrypted message to Bob. The RSA encryption scheme works as follows:

\begin{description}
\item[Key generation] Bob picks randomly two distinct and large prime numbers $p$ and $q$. We denote $n=pq$ and $\phi = (p-1)(q-1)$. Bob then picks a random integer $1< e <\phi$ that is coprime with $\phi$, and computes the inverse $d$ of $e$ modulo $\phi$ ($\gcd(e,\phi)=1$). This inversion can be achieved efficiently using for instance the extended Euclidean algorithm for the greatest common divisor\cite{HW79}. Bob's private key is $\kappa=d$ and his public key is $\pi=(e, n)$

\item[Encryption] Alice obtains Bob's public key $\pi=(e,n)$ from some sort of yellow pages or an RSA public key directory. Alice then writes her message as a sequence of numbers using, for example, our digital alphabet. This string of numbers is subsequently divided into blocks such that each block when viewed as a number $P$ satisfies $P\leq n$. Alice encrypts each $P$ as
\begin{equation}
C = \hat E_\pi(P) =  P^e \bmod n
\end{equation}
and sends the resulting cryptogram to Bob.

\item[Decryption] Receiving the cryptogram $C$, Bob decrypts it by calculating
\begin{equation}
\hat D_\kappa(C) = C^d \bmod n = P
\end{equation}
where the last equality will be proved shortly.
\end{description}

The mathematics behind the RSA is a lovely piece of number theory which goes
back to the XVI century when a French lawyer Pierre de Fermat discovered
that if a prime $p$ and a positive integer $a$ are coprime, then
\begin{equation}
a^{p-1} = 1 \bmod p.
\end{equation}

The cryptogram $C=P^e \bmod n$ is decrypted by $C^d \bmod n = P^{ed} \bmod n$ because $ed = 1 \bmod \phi$, implying the existence of an integer $k$ such that $ed=k\phi+1=k(p-1)(q-1)+1$. If $P \neq 0 \bmod p$, using equation~(9.5) this implies
\begin{equation}
P^{ed} \bmod p = \left(P^{(p-1)}\right)^{k(q-1)}P \bmod p = P \bmod p.
\end{equation}

The above equality holds trivially in the case $P=0\bmod p$. By identical arguments, $P^{ed} \bmod q = P \bmod q$. Since $p$ and $q$ are distinct primes, it follows that
\begin{equation}
P^{ed} \bmod n = P.
\end{equation}

For example, let us suppose that Bob's public key is
$\pi=(e,n)=(179, 571247) $.~\footnote{ Needless to say, number $n$
in this example is too small to guarantee security, do not try this
public key with Bob.} He generated it following the prescription
above choosing $p=773$, $q=739$ and $e=179$. The private key $d$
was obtained by solving $179 d = 1\bmod 772\times 738$ using the
extended Euclidean algorithm which yields $d=515627$. Now if we
want to send Bob encrypted ``SHAKEN NOT STIRRED'' we first use our
digital alphabet to obtain the plaintext which can be written as
the following sequence of six digit numbers
\begin{center}
\begin{tabular}{cccccc}
180700 & 100413 & 261314 & 192618 & 190817 & 170403
\end{tabular}
\end{center}
Then we encipher each block $P_i$ by computing $C_i=P_i^e \bmod n$; \emph{e.g.} the
first block $P_1=180700$ will be eciphered as
\begin{equation}
P_1^e \bmod n = 180700^{179} \bmod 571247 = 141072 = C_1,
\end{equation}
and the whole message is enciphered as:
\begin{center}
\begin{tabular}{cccccc}
141072 & 253510 & 459477 & 266170 & 286377 & 087175
\end{tabular}
\end{center}
The cryptogram $C$ composed of blocks $C_i$ can be send over to Bob. He
can then decrypt each block using his private key $d=515627$, \emph{e.g.} the first
block is decrypted as
\begin{equation}
141072^{515627} \bmod 571247 = 180700 = P_1.
\end{equation}

In order to recover plaintext $P$ from cryptogram $C$, an outsider, who
knows $C$, $n$, and $e$, would have to solve the congruence
\begin{equation}
P^e \bmod n = C,
\end{equation}
for example, in our case,
\begin{equation}
P_1^{179} \bmod 571247 = 141072.
\end{equation}
Solving such an equation is believed to be a hard computational
task for classical computers. So far, no classical algorithm has
been found that computes the solution efficiently when $n$ is a
large integer (say $200$ decimal digits long or more). However, if
we know the prime decomposition of $n$ it is a piece of cake to
figure out the private key $d$: we simply follow the key generation
procedure and solve the congruence $ed = 1\bmod (p-1)(q-1)$. This
can be done efficiently even when $p$ and $q$ are very large. Thus,
in principle, anybody who knows $n$ can find $d$ by factoring $n$.
The security of RSA therefore relies among others on the assumption
that factoring large numbers is computationally difficult. In the
context of classical computation, such difficulty has never been
proved. Worse still, we have seen in Section 8 that there is a
quantum algorithm that factors large number efficiently. This means
that the security of the RSA cryptosystem will be completely
compromised if large-scale quantum computation becomes one day
practical. This way, the advent of quantum computation rules out
public cryptographic schemes commonly used today that are based on
the ``difficulty'' of factoring or the ``difficulty'' of another
mathematical operation called discrete logarithm~\cite{HW79}.

On the other hand, quantum computation provides novel techniques to
generate a shared private key with perfect confidentiality,
regardless the computational power (classical or quantum) of the
adversaries. Such techniques are referred to as \emph{quantum key
distribution} protocols and were proposed independently in the 
United States (S.Wiesner~\cite{Wie83}, C.H.~Bennett and
G.~Brassard~\cite{BB84}) and in Europe (A. Ekert~\cite{Eke91}). 
Discussion on quantum key distribution is
outside the scope of this lecture.

\section{Conditional quantum dynamics}
Quantum gates and quantum networks provide a very convenient language for
building any quantum computer or (which is basically the same) quantum
multiparticle interferometer. But can we build quantum logic gates?

Single qubit quantum gates are regarded as relatively easy to implement. For
example, a typical quantum optical realisation uses atoms as qubits and
controls their states with laser light pulses of carefully selected
frequency, intensity and duration; any prescribed superposition of two
selected atomic states can be prepared this way.

Two-qubit gates are much more difficult to build.

In order to implement two-qubit quantum logic gates it is sufficient, from
the experimental point of view, to induce a conditional dynamics of physical
bits, \emph{i.e.} to perform a unitary transformation on one physical subsystem
conditioned upon the quantum state of another subsystem,
\begin{equation}
U = \left | \, 0 \right \rangle\left \langle 0 \, \right |\otimes U_0 +
\left | \, 1 \right \rangle\left \langle 1 \, \right |\otimes U_1 + \cdots +
\left | \, k \right \rangle\left \langle k \, \right |\otimes U_k,
\end{equation}
where the projectors refer to quantum states of the control subsystem and
the unitary operations $U_i$ are performed on the target subsystem~\cite{BDEJ95}. The simplest non-trivial operation of this sort is probably a
conditional phase shift such as $B(\phi)$ which we used to
implement the quantum Fourier transform and the quantum
controlled-\textsc{not} (or
\textsc{xor}) gate.

Let us illustrate the notion of the conditional quantum dynamics
with a simple example. Consider two
qubits, \emph{e.g.} two spins, atoms, single-electron quantum dots, which
are coupled via a
$\sigma^{(1)}_z\sigma^{(2)}_z$ interaction (\emph{e.g.} a dipole-dipole
interaction):
\setlength{\unitlength}{0.017in}
\begin{center}
\begin{picture}(190,80)

\put(20,50){\oval(40,50)}
\put(10,35){\line(1,0){15}}
\put(27,36){${\ket{0}}$}
\put(10,64){\line(1,0){15}}
\put(27,60){${\ket{1}}$}
\put(0,12){$\hat H_1 = \hbar\omega_1 \sigma_z^{(1)}$}

\put(120,50){\vector(1,0){17}}
\put(70,48) {$\hat V=\hbar \Omega \sigma_z^{(1)}\sigma_z^{(2)}$}
\put(60,50){\vector(-1,0){17}}

\put(160,50){\oval(40,50)}
\put(150,40){\line(1,0){15}}
\put(167,36){${\ket{0}}$}
\put(150,59){\line(1,0){15}}
\put(167,60){${\ket{1}}$}
\put(140,12){$\hat H_2 = \hbar\omega_2 \sigma_z^{(2)}$}
\end{picture}
\end{center}
The first qubit, with resonant frequency
$\omega_1$, will act as the control qubit and the second one, with
resonant frequency $\omega_2$, as the target qubit. Due to the coupling
$\hat{V}$ the resonant
frequency for transitions between the states $\left | \, 0 \right
\rangle$ and $\left | \, 1 \right \rangle$ of one qubit
\emph{depends on the neighbour's state}. The resonant frequency for
the first qubit becomes $
\omega_1\pm\Omega$ depending on whether the second qubit is in state $\left
| \, 0 \right \rangle$ or $\left | \, 1 \right \rangle$. Similarly
the second qubit's resonant frequency becomes $\omega_2\pm\Omega$,
depending on the state of the first qubit. Thus a $\pi$-pulse at
frequency $
\omega_2+\Omega$ causes the transition $\left | \, 0 \right \rangle
\leftrightarrow \left | \, 1 \right \rangle$ in the second qubit only if the
first qubit is in $\left | \, 1 \right \rangle$ state. This way we can
implement the quantum controlled-\textsc{not} gate.

\section{Decoherence and recoherence}
Thus in principle we know how to build a quantum computer; we can
start with simple quantum logic gates and try to integrate them
together into quantum networks. However, if we keep on putting
quantum gates together into networks we will quickly run into some
serious practical problems. The more interacting qubits are
involved the harder it tends to be to engineer the interaction that
would display the quantum interference. Apart from the technical
difficulties of working at single-atom and single-photon scales,
one of the most important problems is that of preventing the
surrounding environment from learning about which computational
path was taken in the multi-particle interferometer. This ``welcher
Weg'' information can destroy the interference and the power of
quantum computing.

Consider the following qubit-environment interaction, known as {\em
decoherence}\cite{Zur91},
\begin{equation}
|0,m\rangle \mapsto |0,m_{0}\rangle, \qquad
|1,m\rangle \mapsto |1,m_{1}\rangle,
\end{equation}
where $|m\rangle $ is the initial state and $|m_{0}\rangle $,
$|m_{1}\rangle $ are the two final states of the environment. This
is basically a measurement performed by the environment on a qubit.
Suppose that in our single qubit interference experiment
(see Eqs.~(\ref{eqinterfere}))
a qubit in between the two Hadamard transformation is
``watched'' by the environment which learns whether the qubit is in
state $|0\rangle $ or $|1\rangle .$ The evolution of the qubit and
the environment after the first Hadamard and the phase gate is
described by the following transformation,
\begin{equation}
\left| \,0\right\rangle \left| \,m\right\rangle \stackrel{H}{\mapsto
}\frac{1}{\sqrt{2}}\left( \left| \,0\right\rangle +\left| \,1\right\rangle
\right) \left| \,m\right\rangle \stackrel{\phi }{\mapsto }\frac{1}{
\sqrt{2}}(e^{i\phi /2}\left| \,0\right\rangle +e^{-i\phi /2}\left|
\,1\right\rangle )\left| \,m\right\rangle.
\end{equation}
We write the decoherence action as
\begin{equation}
\frac{1}{\sqrt{2}}(e^{i\frac{\phi }{2}}\left| \,0\right\rangle +e^{-i\frac{
\phi }{2}}\left| \,1\right\rangle )\left| \,m\right\rangle \stackrel{}{
\mapsto }\frac{1}{\sqrt{2}}(e^{i\frac{\phi }{2}}\left|
\,0\right\rangle \left| \,m_{0}\right\rangle +e^{-i\frac{\phi }{2}}\left|
\,1\right\rangle \left| \,m_{1}\right\rangle ).
\end{equation}
The final Hadamard gate generates the output state
\begin{eqnarray}
\lefteqn{\frac{1}{\sqrt{2}}(e^{i\frac{\phi }{2}}\left| \,0\right\rangle \left|
\,m_{0}\right\rangle +e^{-i\frac{\phi }{2}}\left| \,1\right\rangle \left|
\,m_{1}\right\rangle )}\\
&\stackrel{H}{\mapsto }&\frac{1}{2}\ket{0}\left( e^{i
\frac{\phi }{2}\,}\left| \,m_{0}\right\rangle +e^{-i\frac{\phi }{2}}\left|
\,m_{1}\right\rangle \right)\nonumber\\
&+& \frac{1}{2} \ket{1} \left( e^{i
\frac{\phi }{2}\,}\left| \,m_{0}\right\rangle -e^{-i\frac{\phi }{2}}\left|
\,m_{1}\right\rangle \right).
\end{eqnarray}

Taking $\left| \,m_{0}\right\rangle $ and $\left|
\,m_{1}\right\rangle $ to be normalised and $\langle m_{0}\left|
\,m_{1}\right\rangle $ to be real we obtain the probabilities $P_{0}$ and $
P_{1}$,
\begin{eqnarray}
P_{0} &=&\frac{1}{2}\left( 1+\langle m_{0}\left| \,m_{1}\right\rangle \cos
\phi \right) ,  \label{decoh} \\
P_{1} &=&\frac{1}{2}\left( 1-\langle m_{0}\left| \,m_{1}\right\rangle \cos
\phi \right) .
\end{eqnarray}

It is instructive to see the effect of decoherence on the qubit
alone when its state is written in terms as a density operator. The
decoherence interaction entangles qubits with the environment,
\begin{equation}
\left( \alpha |\,0\rangle +\beta |1\rangle \right) |m\rangle \mapsto
\alpha |\,0\rangle |m_{0}\rangle +\beta |\,1\rangle |m_{1}\rangle .
\label{deco}
\end{equation}
Rewriting in terms of density operators and tracing over the environment's
Hilbert space on the both sides, we obtain
\begin{equation} \label{eqdecohere}
\left(
\begin{array}{cc}
\left| \alpha \right| ^{2} & \alpha \beta ^{\ast } \\
\alpha ^{\ast }\beta & \left| \beta \right| ^{2}
\end{array}
\right) \mapsto \left(
\begin{array}{cc}
\left| \alpha \right| ^{2} & \alpha \beta ^{\ast }\langle
m_{0}|\,m_{1}\rangle \\
\alpha ^{\ast }\beta \langle m_{1}|\,m_{0}\rangle & \left| \beta \right| ^{2}
\end{array}
\right) .
\end{equation}
The off-diagonal elements, originally called by atomic physicists
coherences, vanish as $\langle m_{1}|\,m_{0}\rangle
\mapsto 0,$ that is why this particular interaction with the
environment is called decoherence.

How does decoherence affect, for example, Deutsch's algorithm?
Substituting $ 0$ or $\pi $ for $\phi $ in Eq.(\ref{decoh}) we see
that we obtain the correct answer only with some probability, which
is
\begin{equation}
\frac{1+\langle m_{0}|\,m_{1}\rangle }{2}.
\end{equation}
If $\langle m_{0}|\,m_{1}\rangle =0$, the perfect decoherence case,
then the network outputs $0$ or $1$ with equal probabilities, \emph{i.e.}
it is useless as a computing device. It is clear that we want to
avoid decoherence, or at least diminish its impact on our computing
device.

In general when we analyse {\em physically realisable} computations
we have to consider errors which are due to the
computer-environment coupling and from the computational complexity
point of view we need to assess how these errors scale with the
input size $n$. If the probability of an error in a single run,
$\delta (n)$, grows exponentially with $n$, \emph{i.e.} if $\delta
(n)= 1-A\exp (-\alpha n)$, where $A$ and $\alpha$ are positive
constants, then the randomised algorithm cannot technically be
regarded as efficient any more regardless of how weak the coupling
to the environment may be. Unfortunately, the computer-environment
interaction leads to just such an unwelcome exponential increase of
the error rate with the input size. To see this consider a register
of size $n$ and assume that each qubit decoheres separately,
\begin{eqnarray}
\lefteqn{\ket{x}\ket{M}=\ket{x_{n-1}\ldots x_1x_0}\ket{m}\ldots\ket{m}\ket{m}}\nonumber\\
&\mapsto& \ket{x_{n-1}\ldots x_1x_0}\ket{m_{x_{n-1}}}...\ket{m_{x_1}}\ket{m_{x_0}}=\ket{x}\ket{M_x},
\end{eqnarray}
where $x_i\in\{0,1\}$. Then a superposition
$\alpha\ket{x}+\beta\ket{y}$ evolves as
\begin{equation}
(\alpha\ket{x}+\beta\ket{y})\ket{M}\mapsto
\alpha\ket{x}\ket{M_x}+\beta\ket{y}\ket{M_y},
\end{equation}
but now the scalar product $\bra{M_x} M_y\rangle$ which reduces the
off-diagonal elements of the density operator of the whole register
and which affects the probabilities in the interference experiment
is given by
\begin{equation}
\bra{M_x} M_y\rangle =\bra{m_{x_0}} m_{y_0}\rangle
\bra{m_{x_1}} m_{y_1}\rangle ...\bra{m_{x_{n-1}}} m_{y_{n-1}}\rangle
\end{equation}
which is of the order of
\begin{equation}
\bra{M_x} M_y\rangle=\bra{m_0} m_1\rangle^{H(x,y)},
\end{equation}
where $H(x,y)$ is the Hamming distance between $x$ and $y$, \emph{i.e.}
the number of binary places in which $x$ and $y$ differ (\emph{e.g.} the
Hamming distance between $101101$ and $111101$ is $1$ because the
two binary string differ only in the second binary place). Hence
there are some coherences which disappear as $\bra{m_0}
m_1\rangle^n$ and therefore in some interference experiments the
probability of error may grow exponentially with $n$.

It is clear that for quantum computation of any reasonable length
to ever be physically feasible it will be necessary to incorporate
some efficiently realisable stabilisation scheme to combat the
effects of decoherence. Deutsch was the first one to discuss this
problem. During the Rank Prize Funds Mini--Symposium on Quantum
Communication and Cryptography, Broadway, England in 1993 he
proposed `recoherence' based on a symmetrisation procedure (for
details see~\cite{BBDEJM}). The basic idea is as follows. Suppose
we have a quantum system, we prepare it in some initial state
$\ket{\Psi_i}$ and we want to implement a prescribed unitary
evolution $\ket{\Psi (t)}$ or just preserve $\ket{\Psi_i}$ for some
period of time $t$. Now, suppose that instead of a single system we
can prepare $R$ copies of $\ket{\Psi_i}$ and subsequently we can
project the state of the combined system into the symmetric
subspace \emph{i.e.} the subspace containing all states which are
invariant under any permutation of the sub-systems. The claim is
that frequent projections into the symmetric subspace will reduce
errors induced by the environment. The intuition behind this
concept is based on the observation that a prescribed error-free
storage or evolution of the $R$ independent copies starts in the
symmetric sub-space and should remain in that sub-space. Therefore,
since the error-free component of any state always lies in the
symmetric subspace, upon successful projection it will be unchanged
and part of the error will have been removed. Note however that the
projected state is generally not error--free since the symmetric
subspace contains states which are not of the simple product form
$\ket{\Psi}\ket{\Psi}\ldots\ket{\Psi}$. Nevertheless it has been
shown that the error probability will be suppressed by a factor of
$1/R$ \cite{BBDEJM}.

More recently projections on symmetric subspaces were replaced by
more complicated projections on carefully selected subspaces. These
projections, proposed by Shor~\cite{Sho95}, Calderbank and
Shor~\cite{CS96}, Steane~\cite{Ste96} and
others~\cite{EM96,LMPZ96,Got96,CRSS97,KL96}, are constructed on the
basis of classical error-correcting methods but represent
intrinsically new quantum error-correction and stabilisation
schemes; they are the subject of much current study.

Let us illustrate the main idea of recoherence by describing a
simple method for protecting an unknown state of a single qubit in
a noisy quantum register. Consider the following scenario: we want
to store in a computer memory one qubit in an unknown quantum state
of the form $\ket{\phi}=\alpha\ket{0}+\beta\ket{1}$ and we know
that any single qubit which is stored in a register undergoes a
decoherence type entanglement with an environment described by
Eq.(\ref{deco}). To see how the state of the qubit is affected by
the environment, we calculate the fidelity of the decohered state
at time $t$ with respect to the initial state $\ket{\phi}$
\begin{equation}
F(t)=\bra{\phi}\rho(t)\ket{\phi},
\end{equation}
where $\rho(t)$ is given by Eq.~(\ref{eqdecohere}). It follows that
\begin{equation}
F(t)=|\alpha|^4 + |\beta|^4+2|\alpha|^2 |\beta|^2
\mbox{Re}[\bra{m_0(t)}m_1(t)
\rangle]\;.
\end{equation}

The expression above depends on the initial state $\ket{\phi}$ and
clearly indicates that some states are more vulnerable to decoherence
than others.  In order to get rid of this dependence we consider the
average fidelity, calculated under the assumption that any initial
state $\ket{\phi}$ is equally probable.  Taking into account the
normalisation constraint the average fidelity is given by
\begin{equation}
\bar F(t)=\int_0^1 F(t) \;d\; |\alpha|^2=
\frac{1}{3}(2+\mbox{Re}[\bra{m_0(t)}m_1(t)\rangle])\;.
\end{equation}
If we assume an exponential-type decoherence, where
$\bra{m_0(t)}m_1(t)\rangle=e^{-\gamma t}$, the average fidelity
takes the simple form
\begin{equation}
\bar F(t)=\frac{1}{3}(2+e^{-\gamma t})\;.
\label{fiwec}
\end{equation}
In particular, for times much shorter than the decoherence time
$t_{d}=1/\gamma$, the above fidelity can be approximated as
\begin{equation}
\bar F(t)\simeq1-\frac{1}{3}\gamma t +O(\gamma^2 t^2)\;.
\label{afwec}
\end{equation}

Let us now show how to improve the average fidelity by quantum encoding.
Before we place the qubit in the memory register we {\em encode} it:
we can add two qubits, initially both in state $\ket{0}$, to the
original qubit and then perform an encoding unitary transformation
\begin{eqnarray}
\ket{000}&\mapsto &\ket{\bar 0\bar 0\bar
0}=(\ket{0}+\ket{1})(\ket{0}+\ket{1})(\ket{0}+\ket{1}),\\
\ket{100}&\mapsto &\ket{\bar 1\bar 1\bar
1}=(\ket{0}-\ket{1})(\ket{0}-\ket{1})(\ket{0}-\ket{1}),
\end{eqnarray}
generating state $\alpha\ket{\bar 0\bar 0\bar 0}+\beta\ket{\bar
1\bar 1\bar 1}$, where $\ket{\bar 0} = \ket{0}+\ket{1}$ and
$\ket{\bar 1}
=\ket{0}-\ket{1}$.  Now, suppose that only the second stored qubit was
affected by decoherence and became entangled with the environment:
\begin{eqnarray}
&& \alpha
(\ket{0}+\ket{1})(\ket{0}\ket{m_0}+\ket{1}\ket{m_1})(\ket{0}+\ket{1})
+\nonumber\\ && \beta
(\ket{0}-\ket{1})(\ket{0}\ket{m_0}-\ket{1}\ket{m_1})(\ket{0}-\ket{1}),
\end{eqnarray}
which can also be  written as
\begin{equation}
(\alpha \ket{\bar 0\bar 0\bar 0} +\beta\ket{\bar 1\bar 1\bar
1})(\ket{m_0} +
\ket{m_1})+
(\alpha \ket{\bar 0\bar 1\bar 0} +\beta\ket{\bar 1\bar 0\bar
1})(\ket{m_0}
- \ket{m_1}).
\end{equation}

The decoding unitary transformation can be constructed using a couple
of quantum controlled-NOT gates and the Toffoli gate, thus completing
the error-correcting network:
\setlength{\unitlength}{0.016 in}
\begin{center}
\begin{picture}(220,100)(20,30)
\put(20,40){ENCODING}
\put(75,40){DECOHERENCE AREA}
\put(190,40){DECODING}

\put(0,54){\makebox(20,12){$\ket{0}$}}
\put(0,74){\makebox(20,12){$\ket{0}$}}
\put(10,111){\makebox(20,12){$\alpha\ket{0}+\beta\ket{1}$}}

\put(246,54){\makebox(20,12){$\ket{x_2}$}}
\put(246,74){\makebox(20,12){$\ket{x_1}$}}
\put(230,111){\makebox(20,12){$\alpha\ket{0}+\beta\ket{1}$}}

\put(20,110){\vector(0,-1){10}}
\put(244,100){\vector(0,1){10}}

\put(20,60){\line(1,0){30}}
\put(20,80){\line(1,0){30}}
\put(20,100){\line(1,0){30}}

\put(30,100){\circle*{4}}
\put(30,100){\line(0,-1){24}}
\put(30,80){\circle{8}}
\put(40,100){\circle*{4}}
\put(40,100){\line(0,-1){44}}
\put(40,60){\circle{8}}

\put(50,94){\framebox(12,12){$H$}}
\put(50,74){\framebox(12,12){$H$}}
\put(50,54){\framebox(12,12){$H$}}

\put(62,60){\line(1,0){10}}
\put(62,80){\line(1,0){10}}
\put(62,100){\line(1,0){10}}

\put(72,50){\framebox (100,60)
{{$\alpha\ket{\bar 0\bar 0\bar 0} + \beta\ket{\bar 1\bar 1\bar 1}$}}}
\put(120,100){\makebox (4,2) {Encoded State}}
\put(172,60){\line(1,0){10}}
\put(172,80){\line(1,0){10}}
\put(172,100){\line(1,0){10}}

\put(182,94){\framebox(12,12){$H$}}
\put(182,74){\framebox(12,12){$H$}}
\put(182,54){\framebox(12,12){$H$}}

\put(194,60){\line(1,0){50}}
\put(194,80){\line(1,0){50}}
\put(194,100){\line(1,0){50}}

\put(204,100){\circle*{4}}
\put(204,100){\line(0,-1){24}}
\put(204,80){\circle{8}}
\put(214,100){\circle*{4}}
\put(214,100){\line(0,-1){44}}
\put(214,60){\circle{8}}
\put(226,100){\circle{8}}
\put(226,60){\line(0,1){44}}
\put(226,60){\circle*{4}}
\put(226,80){\circle*{4}}

\end{picture}
\end{center}

Careful inspection of the network
shows that any single phase-flip $\ket{\bar 0}
\leftrightarrow \ket{\bar 1}$ will be corrected and the environment
will be effectively disentangled from the qubits. In our particular
case we obtain
\begin{equation}
(\alpha\ket{0}+\beta\ket{1})\;[\ket{00}(\ket{m_0}+\ket{m_1})+
\ket{10}(\ket{m_0}-\ket{m_1})].
\end{equation}
The two auxiliary outputs carry information about the error syndrome -
00 means no error, 01 means the phase-flip occurred in the third
qubit, 10 means the phase-flip in the second qubit and 11 signals the
phase flip in the first qubit.

Thus if only one qubit in the encoded triplet decoheres we can recover
the original state perfectly.  In reality all three qubits decohere
simultaneously and, as the result, only partial recovery of the
original state is possible. In this case lengthy but straightforward
calculations show that the average fidelity of the reconstructed state
after the decoding operation for an exponential-type decoherence is
\begin{equation}
\bar F_{\mbox{ec}}(t)=\frac{1}{6}[4+3 e^{-\gamma t}-e^{-3\gamma t}]\;.
\label{fidec}
\end{equation}
For short times this can be written as
\begin{equation}
\bar F_{\mbox{ec}}(t)\simeq 1-\frac{1}{2}\gamma^2 t^2 +O(\gamma^3t^3).
\end{equation}
Comparing Eq.~(\ref{fiwec}) with Eq.~(\ref{fidec}), we can easily see
that for all times $t$,
\begin{equation}
\bar F_{\mbox{ec}}(t)\ge \bar F(t).
\end{equation}

This is the essence of recoherence via encoding and decoding.
There is much more to say (and write) about quantum codes and the
reader should be warned that we have barely scratched the surface of
the current activities in quantum error correction, neglecting topics
such as group theoretical ways of constructing good quantum
codes~\cite{Got96,CRSS97}, concatenated codes~\cite{KL96}, quantum
fault tolerant computation~\cite{fault} and many others.

\section{Concluding remarks}
Research in quantum computation and in its all possible variations
has become vigorously active and any comprehensive review of the
field must be obsolete as soon as it is written. Here we have
decided to provide only some very basic knowledge, hoping that this
will serve as a good starting point to enter the field. Many
interesting papers in these and many related areas can be found at
the Los Alamos National Laboratory e-print archive
(http://xxx.lanl.gov/archive/quant-ph) and on the web site of the
Center for Quantum Computation (www.qubit.org).

\section{Acknowledgments}
This work was supported in part by the European TMR Research
Network ERP-4061PL95-1412, The Royal Society London, and Elsag plc.
PH acknowledges the support of the Rhodes Trust.

\end{document}